\documentclass[reprint,aps,a4paper,showpacs,prc,floatfix,twocolumn,amsmath,amssymb]{revtex4-1}
\usepackage{amsmath,amssymb,amsfonts}

\usepackage{amsmath,amssymb,amsfonts}
\usepackage{graphicx}
\usepackage{color}
\usepackage{floatflt}
\usepackage{longtable}
\usepackage{dcolumn}% Align table columns on decimal point
\usepackage[normalem]{ulem}

\allowdisplaybreaks

\newcommand{\be}{\begin{equation}}
\newcommand{\ee}{\end{equation}}
\newcommand{\bea}{\begin{eqnarray}}
\newcommand{\eea}{\end{eqnarray}}

\begin{document}

%%% \setcounter{page}{1}

%%% \vspace*{0.5 true in}

\title{Strong correlations of neutron star radii with the slopes of nuclear
matter incompressibility and symmetry energy at  saturation
}

\author{N. Alam$^{1,2}$}
\author{B. K. Agrawal$^{1,2}$}
\author {M. Fortin$^3$}
\author{H. Pais$^4$}
\author{C. Provid{\^e}ncia$^4$}
\author{Ad. R. Raduta$^5$}
\author{A. Sulaksono$^6$}

\affiliation{$^1$Saha Institute of Nuclear Physics, Kolkata - 700064, India\\
$^2$Homi Bhabha National Institute, Anushakti Nagar, Mumbai - 400094, India\\ 
$^3$N. Copernicus Astronomical Center, Polish Academy of Science,
Bartycka,18, 00-716 Warszawa, Poland\\
$^4$CFisUC, Department of Physics, University of Coimbra, 3004-516 Coimbra, Portugal\\
$^5$ IFIN-HH, Bucharest-Magurele, POB-MG6, Romania\\
$^6$Departemen Fisika, FMIPA, Universitas Indonesia, Depok, 16424,
Indonesia
}

%\date{\today} 

\begin{abstract} 
We examine the correlations of neutron star radii  with
the nuclear matter incompressibility, symmetry energy, and their
slopes, which are the key parameters of the equation of state (EoS) of
asymmetric nuclear matter.
The neutron star radii and the EoS parameters
are evaluated using  a representative set of 24 
Skyrme-type effective forces and 
18 relativistic mean field  models, and two
microscopic calculations, all describing 2$M_\odot$ neutron stars. Unified
 EoSs for the inner-crust-core
region  have been built for all the phenomenological models, both
  relativistic and non-relativistic.
Our
investigation shows the existence of a strong correlation of the neutron
star radii with the linear combination of the slopes of the nuclear matter
incompressibility and the symmetry energy coefficients at the saturation
density. Such correlations are found to be  almost independent of
the neutron star mass in the range  $0.6\text{-}1.8
  M_{\odot}$. This correlation can be linked to the empirical
relation existing between the star radius and the pressure at a
nucleonic density between one and two times saturation density, and the dependence of the pressure on  the nuclear matter
incompressibility, its slope and the
symmetry energy slope. 
The slopes of the nuclear matter incompressibility and the symmetry
energy coefficients  as estimated from the finite nuclei data yield the
radius of a $1.4M_{\odot}$ neutron star in the range $11.09\text{-}12.86$ km.

% The values of the incompressibility and the symmetry energy slope parameters as
% estimated from the finite nuclei data yield the radius of a $1.4M_{\odot}$
% neutron star in the range  ${\bf R_{1.4}=11.09\text{-}12.86}$ km.

 \end{abstract}


\maketitle

The bulk properties of neutron stars are mainly governed by the
behaviour of the equation of state (EoS) of highly asymmetric
dense matter. The correlations of the various EoS parameters of
asymmetric nuclear matter with the different properties of neutron
star, such as the crust-core transition density and pressure, radii,
maximum mass and cooling rate,  have been studied
\cite{Glendenning86,Horowitz01,Lattimer07,Xu09,vidana09,Ducoin10,Ducoin11,
Cavagnoli11,Gandolfi12,Fattoyev12,Newton13,Fattoyev14,Sotani15,Fortin16,Dutra16}.
The crust-core transition density is strongly correlated with the slope
of the symmetry energy, $L_0$, at saturation density ($\rho_0\sim 0.16$
fm$^{-3}$) \cite{vidana09,Ducoin10,Newton13}. However, the transition pressure
is found to be strongly correlated with a linear combination of the
slope and curvature of the symmetry energy at the sub-saturation
density ($\rho=0.1$ fm$^{-3}$) \cite{Ducoin11,Newton13,Fattoyev14}.
The simultaneous determination of mass and radius of low-mass neutron
stars can better constrain the product of nuclear matter incompressibility
and symmetry energy slope parameter \cite{Sotani15}.

The correlations of the neutron star radii of different masses with the
EoS parameters have been investigated extensively. The covariance analysis,
based on a single model, suggests the existence  of strong correlations of
the radii of low-mass neutron stars ($M_{\rm NS}\sim 0.6\text{-}1.2M_{\odot}$)
with the symmetry energy slope parameter $L_0$ \cite{Fattoyev12}, the
correlations becoming weaker with the increase of the neutron star mass.
Similar analysis for the correlations of the radii with the symmetry
energy slope over a wider range of densities was performed for two
different models, having different behaviours on the density dependence of
the symmetry energy, and such correlations were found to be model dependent
\cite{Fattoyev14}. Recently, correlations of neutron star radii with the
symmetry energy slope parameter and the nuclear matter incompressibility
coefficient have been examined using a  large set of unified EoSs,
based on Skyrme-type effective forces and relativistic mean field
(RMF) models \cite{Fortin16}. The dependence of correlations on the
neutron star mass is qualitatively similar to those obtained within the
covariance analysis, but the correlations are in general somewhat weaker
due to the interference of the other EoS parameters, which were kept
fixed in the later case.  Since the EoS for asymmetric nuclear matter
is mainly governed by the nuclear matter incompressibility, symmetry
energy and their slopes at saturation density, the neutron star radii
may be correlated with the linear combination of these EoS parameters,
rather than each parameter individually, analogous to those found in the
case of the correlation between the transition pressure and the linear combination
of the slope and the curvature of the symmetry energy \cite{Ducoin11}.

In this Rapid Communication, we examine the correlations
of the neutron star radii with the key parameters governing the EoS of
 asymmetric nuclear matter.  These EoS parameters are evaluated
at the nuclear saturation density, using a representative set of RMF
models, a set of Skyrme-type models, and one  microscopic
  calculation using  Brueckner-Hartree-Fock (BHF) with
the Argonne $V_{18}$ force, and a three body force of Urbana type
\cite{Taranto13}, and  a variational approach, in particular the 
Akmal-Pandharipande-Ravenhall (APR) EoS \cite{APR}. All models describe 2
 $M_\odot$ stars.
We demonstrate that the  neutron star radii over a wide range
of masses ($0.6\text{-}1.8M_{\odot}$) are strongly correlated with the
linear combination of the slopes of nuclear matter incompressibility
and symmetry energy coefficients.

The EoS at a given density $\rho$ and asymmetry $\delta$ can be decomposed, to a good approximation, into the EoS for symmetric nuclear matter $e(\rho,0)$, and the density dependent symmetry energy coefficient $S(\rho)$ as
 \begin{equation}
 e(\rho,\delta)=e(\rho,0)+S(\rho)\delta^2 \, ,\\
 \label{eq:eden}
\end{equation}
 where $e(\rho,\delta)$ is the energy per nucleon
at density $\rho=\rho_n+\rho_p$, and  $\delta=(\rho_n-\rho_p)/\rho$ the asymmetry parameter, with $\rho_n$ and $\rho_p$ 
 the neutron and proton densities, respectively. 

 The isoscalar part $e(\rho,0)$ can be expanded as
\begin{small}
\begin{eqnarray}
e(\rho,0)=e(\rho_0)+\frac{K_0}{2}\left(\frac{\rho-\rho_0}{3\rho_0}\right)^2+\frac{Q_0}{6}\left(\frac{\rho-\rho_0}{3\rho_0}\right)^3+\mathcal{O}(4)
\label{eq:E}\hspace{+13pt}
\end{eqnarray}
\end{small}
and the isovector part $S(\rho)$ as
\begin{small}
\begin{eqnarray}
S(\rho)=J_0+L_0\left(\frac{\rho-\rho_0}{3\rho_0}\right)+\frac{K_{sym,0}}{2}\left(\frac{\rho-\rho_0}{3\rho_0}\right)^2+\mathcal{O}(3),\\
\label{eq:S}\hspace{+4pt}
\end{eqnarray}
\end{small}
where $J_0=S(\rho_0)$ is the symmetry energy coefficient.  The incompressibility $K_0$, the skewness coefficient $Q_0$, 
the slope $L_0$, and the curvature $K_{sym,0}$ of the symmetry energy are defined in, e.g.,  Ref. \cite{vidana09}.

The slope of the incompressibility, $M_0$, at saturation density,
and $K_{\tau,0}$  are defined as \cite{Alam15}
\begin{eqnarray}
M_0=Q_0+12K_0 \, ,\\
K_{\tau,0}=K_{sym,0}-6L_0-\frac{Q_0}{K_0}L_0.
\end{eqnarray}

\begin{figure}
 \begin{center}
    \includegraphics[width=0.42\textwidth,angle=0]{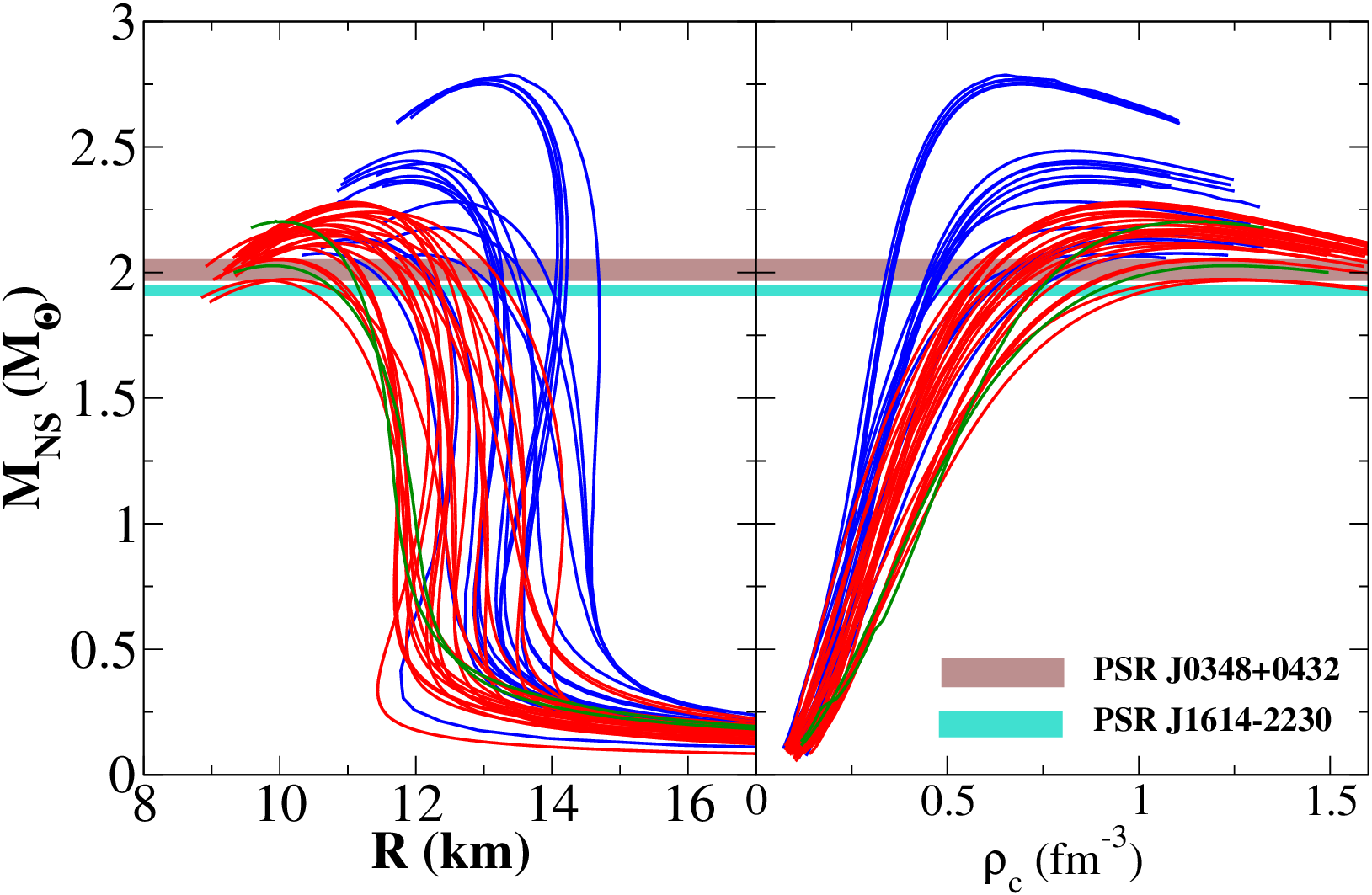}
     \end{center}
  \caption{Neutron star mass in $M_\odot$ as a function of the radius in km (left) 
  and central density in fm$^{-3}$ (right) for a representative set of RMF
    (blue) and Skyrme (red) models, and microscopic (green) calculations. }
 \label{fig1}
\end{figure}

We use a representative set of RMF models, a set of Skyrme-type models, and two microscopic calculations for our
correlation study. The RMF models can be classified 
broadly into two categories: (1) models with nonlinear
self and/or mixed interaction terms and constant coupling strengths
and (2) models with only linear interaction terms but density-dependent 
coupling 
strengths.  The type I models 
used in the present calculations are
BSR2, BSR3, BSR6 \cite{Dhiman07,Agrawal10}, FSU2 \cite{Chen14}, GM1 \cite{Glendenning91}, NL3 \cite{Lalazissis97},
NL$3{\sigma\rho}4$, NL$3{\sigma\rho}6$ \cite{Pais16}, NL$3{\omega \rho}02$ \cite{Horowitz01}, 
NL$3{\omega \rho}03$ \cite{Carriere03}, TM1 \cite{Sugahara94}, and TM1-2 \cite{Providencia13}. The type II models are 
DD2 \cite{Typel10}, DDH$\delta$ \cite{Gaitanos04}, DDH$\delta$Mod \cite{Ducoin11}, DDME1 \cite{Niksic02}, 
DDME2 \cite{Lalazissis05}, and TW \cite{Typel99}. The Skyrme models we use in this work are SKa, SKb \cite{Kohler76},
SkI2, SkI3, SkI4, SkI5 \cite{Reinhard95}, SkI6 \cite{Nazarewicz96}, Sly2, Sly9 \cite{ChabanatPhd},
Sly230a \cite{Chabanat97}, Sly4 \cite{Chabanat98}, SkMP \cite{Bennour89}, SKOp \cite{Reinhard99},
KDE0V1 \cite{Agrawal05}, SK255, SK272 \cite{Agrawal03}, Rs \cite{Friedrich86}, BSk20, BSk21
\cite{Goriely10}, BSk22, BSk23, BSk24, BSk25, and BSk26
\cite{Goriely13}. The microscopic calculations include the BHF EoS from
\cite{Taranto13,Davesne16}, and the APR  EoS is taken from \cite{APR,Steiner05,Ducoin11}. 
The values of the EoS parameters at saturation density show a wide variation across the models. 
 The symmetric nuclear matter properties for these models are presented in Table I of the Supplemental Material.
   We shall mainly focus on the correlations between the neutron star radii and the
various key parameters of the EoSs: $K_0$, $Q_0$, $M_0$, $J_0$, $L_0$, $K_{sym,0}$, and $K_{\tau,0}$,
which are  evaluated at saturation density.

%BSR10, BSR17, FSU \cite{Todd-Rutel05}, IUFSU \cite{Fattoyev10},
%DD2 \cite{Typel10}, DDH$\delta$ \cite{Gaitanos04}, DDH$\delta$Mod \cite{Ducoin11}, DDME1 \cite{Niksic02}, DDME2 \cite{Lalazissis05}

It was shown in Ref. \cite{Fortin16} that non-unified EoSs may
introduce a large uncertainty on the determination of low-mass star
radii, i.e. $M_{\rm NS}\lesssim 1.4 M_\odot$, mainly if the behaviours
of the symmetry energy slope for the EoSs of the inner crust and
core are very different.  
%Therefore, we will only consider an EoS
%with the inner crust and core EoS calculated within the same model \mf{(except for APR and BHF most probably?)}.
For the RMF models, the EoSs for $\beta$-equilibrated matter are built according
to the following procedure. The EoS for the outer crust region is
taken from the work of Baym-Pethick-Sutherland \cite{Baym71}. For the
inner crust region, we use the EoS including the pasta phases, if they
exist, obtained within a Thomas Fermi calculation \cite{Grill14} up
to the crust-core transition density, $\rho_t$.  At the crust-core
transition, the inner crust EoS is matched to the corresponding
homogeneous EoS. The fraction of the particles at a given density
is determined imposing $\beta$-equilibrium and charge neutrality.
%  Consequently these EoS are not fully unified since
%   the model used for the outer crust and the one for the inner crust
%   and the core are not the same. 
The  model used for the outer crust is not the  same as the one used for
the inner crust and the core regions. However, the use of different EoSs for
 the outer crust has been shown to barely affect the radius of a star
  for masses above $1M_{\odot}$ \cite{Fortin16}.
For the Skyrme models, the same functional is used for the crust and
the core. In the crust, a compressible liquid-drop model (CLDM) and a
variational approach, detailed in \cite{Gulminelli2015,Fortin16}, are employed to
describe the nuclei. 
Finally, for the BHF and APR EoSs, the outer and
inner crusts are described by the EoSs \cite{HZD89} and \cite{Negele73},
respectively. 

The mass $M_{\rm NS}$ and radius $R$ of static neutron stars are obtained by 
 solving the Tolman-Oppenheimer-Volkoff
equations \cite{Weinberg72}, using all of these 44 EoSs. 
The mass-radius
relations are plotted in Fig.\,\ref{fig1}
(left panel), where the horizontal strips indicate the masses of the
 two heaviest neutron stars observed so far: PSR J0348+0432 \cite{Antoniadis13} and PSR J1614-2230 \cite{Demorest10, Fonseca16}. 
 For the BSk models, the $M-R$ relations obtained with EoSs based on a simplified CLDM used in this work are close to the ones calculated with a full microscopic model in \cite{Fantina14,Pearson2014}, see in discussion in \cite{Fortin16}.
To facilitate our discussion, we also display the mass as
a function of central density in the right panel of the same figure.
The EoSs give rise to different neutron star properties.
The spread in the maximum mass is $\sim 0.8M_{\odot}$, and the spread
in the radius of neutron star with canonical mass ($1.4M_{\odot}$) is
$\sim3.1$ km.  In Table II of the Supplemental Material, we provide the
maximum masses together with the radii for different neutron star masses,
calculated for all the models used in this study.

 The values of the EoS parameters and neutron star radii, obtained for these models,
 will be used to study the correlations between these quantities. A linear correlation 
 between any two quantities, $a$ and $b$, can be quantitatively
 studied by  the Pearson's correlation coefficient, $C(a,b)$, given by
\begin{equation}
C(a,b)=\frac{\sigma_{ab}}{\sqrt{\sigma_{aa}\sigma_{bb}}}\, ,\\
\label{eq:cc}
\end{equation}
with the covariance, $\sigma_{ab}$, written as
\begin{equation}
\sigma_{ab}=\frac{1}{N_m}\sum_i a_i b_i -\left(\frac{1}{N_m}\sum_i a_i\right
)\left(\frac{1}{N_m}\sum_i b_i\right ) \, ,
\end{equation}
where the index $i$ runs over the number of models $N_m$\cite{Brandt97}.
In what follows, $a_i$ and $b_i$ correspond to the  neutron star radius for
a fixed mass and a EoS parameter, respectively, obtained for the different models.   
A correlation coefficient equal to 1 in absolute value indicates a perfect linear relation between the two quantities that are considered.

\begin{figure}
 \begin{center}
    \includegraphics[width=.45\textwidth,angle=0]{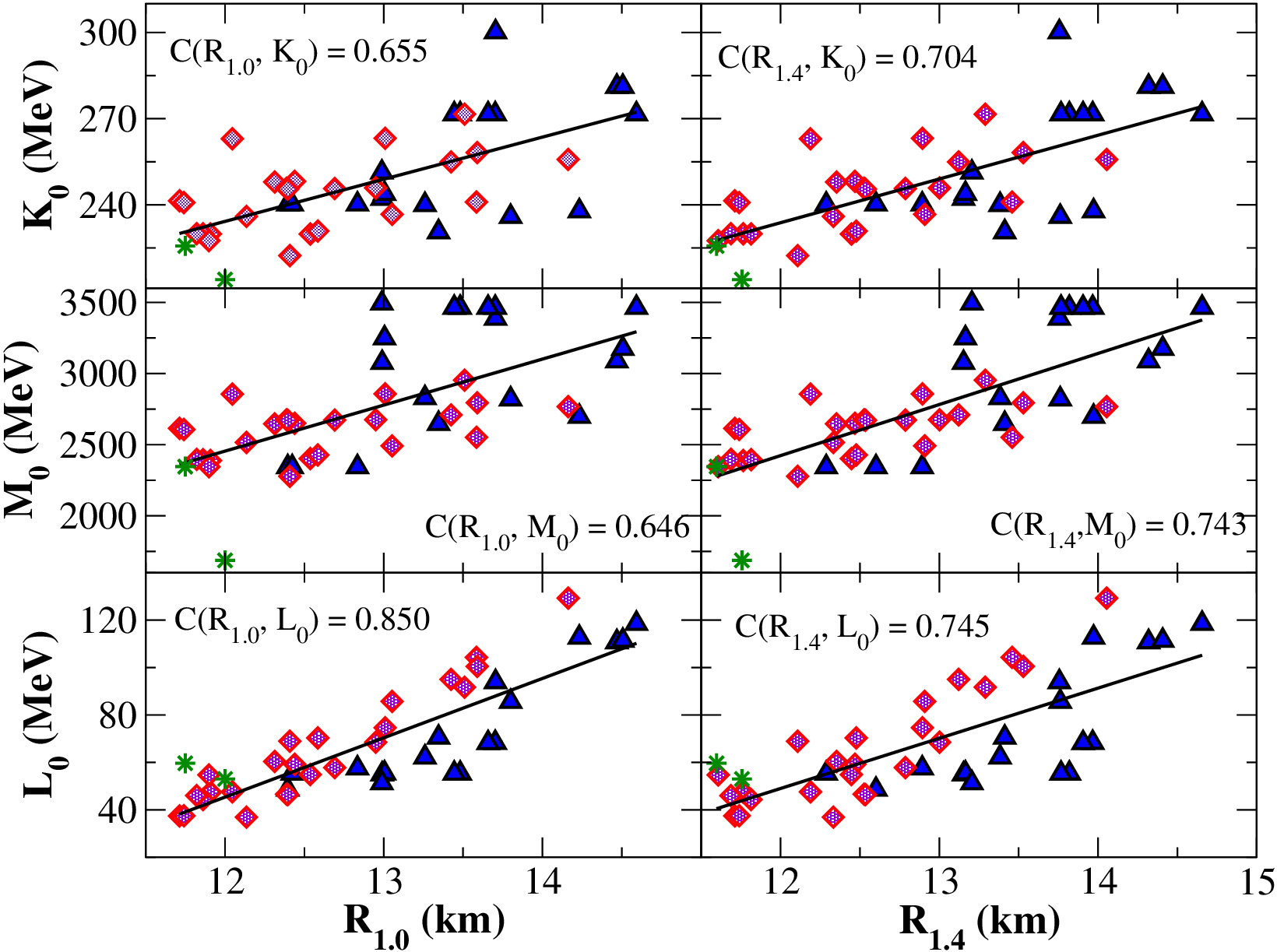}
     \end{center}
  \caption{(Color online) Radii $R_{1.0}$ (left) and $R_{1.4}$ (right) of a $1.0$ and $1.4M_\odot$ neutron star versus the EoS parameters $K_0$, $M_0$ and $L_0$, obtained using our sets of RMF (blue triangles) and Skyrme (red diamonds) models, together with     the BHF and the APR (green stars) calculations.}
 \label{fig2}
\end{figure}

In Fig.\,\ref{fig2}, we plot the radii of  1.0$M_\odot$ and $1.4
M_\odot$ stars,  $R_{1.0}$ and $R_{1.4}$, versus some of these EoS parameters for our representative sets of RMF (blue triangles) and
Skyrme (red diamonds) models, together with the BHF and the APR (green stars) calculations. 
The solid lines in the figures are obtained by linear regression and the correlation coefficients are indicated for each case considered.  
The correlations  between the neutron star
radius and the isoscalar parameters $K_0$ and $M_0$ increase with
the increase of the neutron star mass, however, they
are not significantly strong to make a meaningful prediction.  The
$R_{1.4}\text{-}L_0$ correlation is weaker than the $R_{1.0}\mbox{-}L_0$
correlation, which is opposite to the trend observed for the cases of
$K_0$ and $M_0$. 
In Table\,\ref{tab1}, we list all the correlation coefficients between
the  EoS parameters $K_0$, $M_0$, $Q_0$, $J_0$,  $L_0$, $K_{sym,0} $
and $K_{\tau,0}$, and the radii of neutron stars with different masses.

% The correlation coefficients for the isoscalar (isovector)
%EoS parameters increase (decrease) with the increase of the neutron
%star mass.  
%The neutron star radii are weakly correlated to the EoS
%parameter $K_{sym,0}$ and $Q_0$, and anticorrelated to $K_{\tau,0}$.
The study of the correlations clearly indicates that the radius of low-mass neutron stars
is more sensitive to the isovector EoS parameters ($J_0$ and $L_0$),
but, as the mass of the neutron star increases, the sensitivity to
the isoscalar parameters ($K_0$ and $M_0$) tend to dominate.  A similar
conclusion was drawn in Ref. \cite{Fortin16}, where the behaviour
of the radius of stars with mass $M_{\rm NS}=1.0,\, 1.4,\, 1.8 M_\odot$ for
33 models, including 9 RMF models and 24 Skyrme forces, were plotted as
a function of $K_0$ and $L_0$. The correlation coefficient $C(R_{1.0},
L_0)$ was 0.87, while $C(R_{1.8},L_0)$ was 0.64. The value of $C(R_{1.0},
K_0)$ was 0.63, whereas the values of $C(R_{1.4}, K_0)$ and $C(R_{1.8},
K_0)$ were found to be $\sim 0.66$. These values are quite in agreement with the values we are finding in this work, especially for the correlation coefficients of the low-mass neutron star radii.

 \setlength{\tabcolsep}{4.8pt}
\begin{table}[h!]
\caption{\label{tab1} Correlation coefficients between the neutron star
radii and the different EoS parameters obtained for a
representative set of RMF, Skyrme and microscopic
calculations. The EoS parameters are the nuclear matter incompressibility
coefficient  $K_0$,
its skewness $Q_0$, and slope $M_0$, the symmetry energy coefficient
$J_0$, its slope $L_0$, and curvature
$K_{\rm sym,0}$,  and the parameter $K_{\tau, 0}$, calculated at the saturation density. 
$R_x$ denotes the neutron star radius for a given mass $x$ in units of $M_\odot$.}
 \begin{center}
\begin{tabular}{c|ccccccc}
   & $K_0$  & $Q_0$ & $M_0$ & $J_0$ & $L_0$ & $K_{sym,0}$ & $K_{\tau,0}$\\
\hline 
$R_{0.6}$ & 0.565  &   0.383  &   0.548  &   0.815  &   0.887   &  0.581  &  $ -0.809$ \\
$R_{0.8}$ & 0.617  &   0.416  &   0.597  &   0.743  &   0.881   &  0.658  &  $ -0.775$ \\
$R_{1.0}$ & 0.655  &   0.461  &   0.646  &   0.680  &   0.850   &  0.698  &  $ -0.743$ \\
$R_{1.2}$ & 0.684  &   0.514  &   0.695  &   0.621  &   0.803   &  0.714  &  $ -0.716$ \\
$R_{1.4}$ & 0.704  &   0.571  &   0.743  &   0.562  &   0.745   &  0.711  &  $ -0.689$ \\
$R_{1.6}$ & 0.718  &   0.628  &   0.787  &   0.502  &   0.674   &  0.691  &  $ -0.661$ \\
$R_{1.8}$ & 0.725  &   0.686  &   0.828  &   0.438  &   0.590   &  0.653  &  $ -0.630$ \\
 \hline
\end{tabular}
\end{center}
\end{table}

\begin{figure}
 \begin{center}
    \includegraphics[width=.48\textwidth,angle=0]{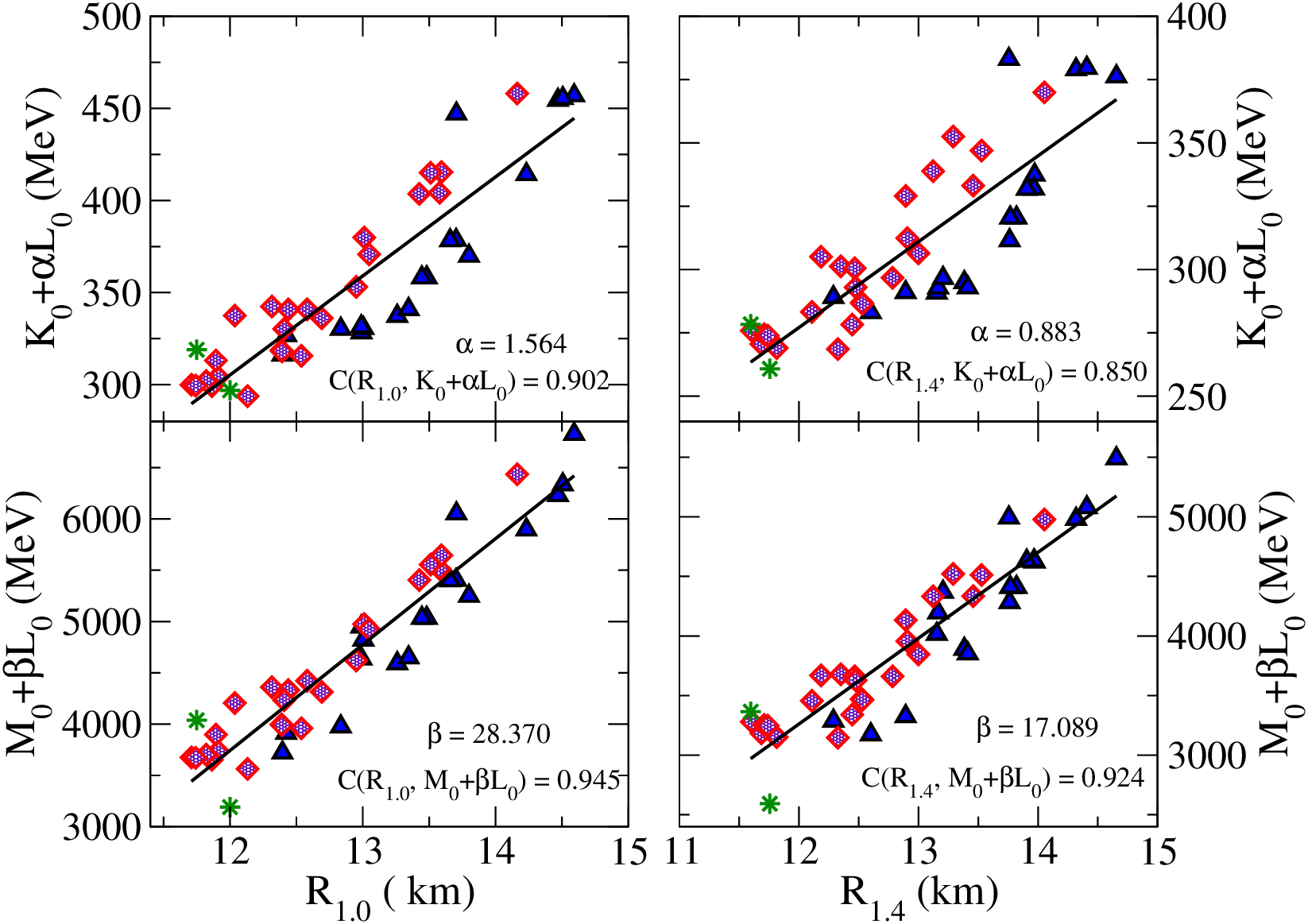}
     \end{center}
  \caption{(Color online) Neutron star radii $R_{1.0}$ (left) and $R_{1.4}$ 
  (right) versus the linear correlations $K_0+\alpha L_0$ (top)  and $M_0+\beta L_0$ (bottom),
  using a set of RMF (blue triangles), Skyrme (red diamonds), and BHF+APR (green stars) calculations.}
 \label{fig3}
\end{figure}
Next we look into the correlations of neutron star radii with
selected combinations of isoscalar and isovector EoS parameters.
In Fig.\,\ref{fig3}, we plot the neutron star radii for $M_{\rm NS}=1.0$
$M_\odot$  (left) and 1.4 $M_\odot$ (right) as a function of the linear
combinations, $K_0+\alpha L_0$ (top), and $M_0+\beta L_0$ (bottom).
We can see that the neutron star radii are better correlated with these
combinations, than with  the  each of the parameter separately,
as seen in Fig.\,\ref{fig2}. Further, the strongest correlations occur
between the neutron star radii and the linear combination $M_0+\beta L_0$
. In Table\,\ref{tab2}, we list again all the correlation coefficients
of neutron star radii with $K_0+\alpha L_0$ and $M_0+\beta L_0$ , for
different neutron star masses.  The values of $\alpha$  and $\beta$,
also listed,  are obtained in such a way that the correlations of these
quantities with the neutron star radii are maximum.

\setlength{\tabcolsep}{6pt}
\begin{table}[h!]
\caption{\label{tab2}The correlation coefficients of neutron star
radii with $K_0+\alpha L_0$ and $M_0+\beta L_0$, along with the 
values of $\alpha$ and $\beta$.}
 \begin{center}
\begin{tabular}{c|cc|cc}
%\hline
%\hline

&  \multicolumn{2}{c}{$K_0+\alpha L_0$} &\multicolumn{2}{|c}{$M_0+\beta L_0$}\\
\hline
          &$\alpha$ & Corr. Coeff.&$\beta$& Corr. Coeff.\\ \cline{2-5}
$R_{0.6}$ &    2.970& 0.905   &  43.115  &   0.936 \\
$R_{0.8}$ &    2.111& 0.914   &  35.575  &   0.949 \\
$R_{1.0}$ &    1.564& 0.902   &  28.370  &   0.945 \\
$R_{1.2}$ &    1.177& 0.879   &  22.189  &   0.935 \\
$R_{1.4}$ &    0.883& 0.850   &  17.089  &   0.924 \\
$R_{1.6}$ &    0.643& 0.817   &  12.781  &   0.913 \\
$R_{1.8}$ &    0.432& 0.782   &   8.970  &   0.903 \\

 \hline
\end{tabular}
\end{center}
\end{table}

\begin{figure}
 \begin{center}
    \includegraphics[width=.38\textwidth,angle=0]{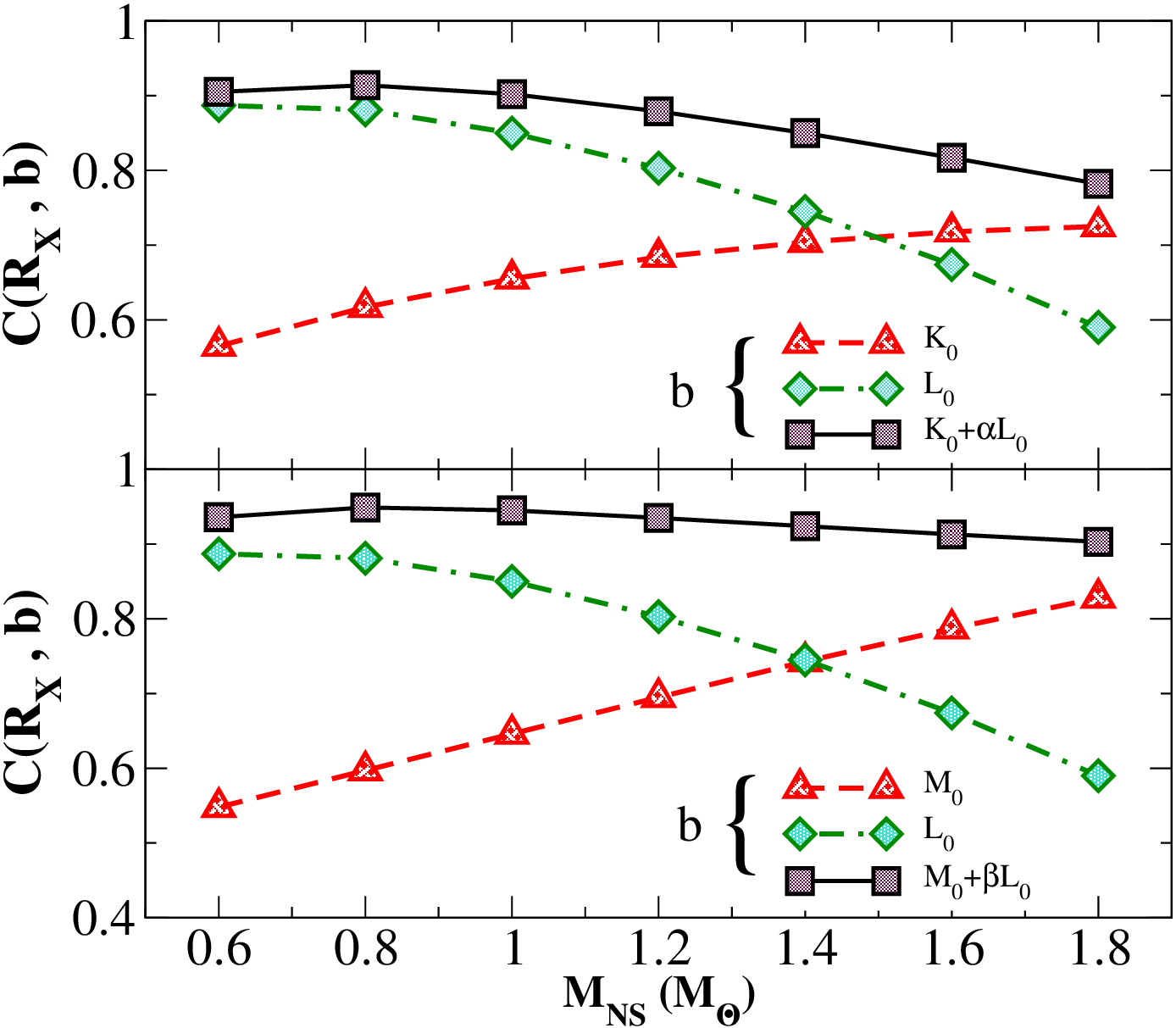}
     \end{center}
  \caption{(Color online) Correlation coefficients between the neutron star radii
 and several EoS parameters as a function of the neutron star mass. The EoS
 parameters '$b$' denote $K_0$, $L_0$, and the linear combination $K_0+\alpha L_0$ 
in the top panel, and $M_0$, $L_0$, and $M_0+\beta L_0$ in the bottom panel. }
 \label{fig4}
\end{figure}

In the top panel of Fig.\,\ref{fig4},
we show the variation of the correlation coefficients of
neutron star radii with $K_0$, $L_0$, and $K_0+\alpha L_0$, as a function
of the mass of the star.  The correlation of neutron star radii with
$K_0+\alpha L_0$ is better than those with $K_0$ and $L_0$ individually.
However, for $M_{\rm NS}\gtrsim1.0M_{\odot}$, the correlations of neutron
star radii decrease gradually with the increase of the neutron star
mass, even considering  $K_0+\alpha L_0$ .  
In the bottom panel of Fig.\,\ref{fig4}, we repeat the same exercise
 for $M_0$, $L_0$, and  $M_0+\beta L_0$. Again,  contrary to
the individual parameters $M_0$ and $L_0$, the neutron star radii
are strongly correlated with $M_0+\beta L_0$ over a wide range of
neutron star masses ($0.6\mbox{ - }1.8M_{\odot}$).

In order to interpret the correlations obtained, we consider the
dependence of the pressure on the isoscalar coefficients, $K_0, \,
M_0,\, Q_0$,  and on the slope of the symmetry energy, $L_0$. Taking the
expansions given in (\ref{eq:E}) and  (\ref{eq:S}), the pressure is
given by 
\begin{equation}
P=\frac{\rho_0x^2}{3}\left[ \frac{K_0}{3} (x-1) +
  \frac{Q_0}{18} (x-1)^2+ L_0 \delta^2\right]\, ,
\label{KL}
\end{equation}
with $x=\rho/\rho_0$, or expressing $Q_0$ in terms of $M_0$ and $K_0$, by
\begin{equation}
P=\frac{\rho_0x^2}{3}\left[ K_0 (x-1)\left(1-\frac{2x}{3}\right) +
  \frac{M_0}{18} (x-1)^2+ L_0 \delta^2\right].
\label{ML}
\end{equation}
These two equations and the empirical relation $R \propto P^{1/4}$,
identified in Ref. \cite{Lattimer01,Lattimer07}, where $R$ is the star
radius and $P$  the pressure, calculated for some fiducial density, $\sim
1-2\rho_0$, allow an interpretation of the above correlations of $R$ with
$K_0+\alpha  L_0$ and $ M_0+\beta L_0$.  

In the following, we present
some arguments that explain the correlations: a) if $\rho=\rho_0$,
only the $L_0$ term survives and this may explain why the radius of
low-mass neutron stars is well correlated with $L_0$; b) from eq. (\ref{ML}),
it is seen that, for $\rho=1.5\rho_0$, the pressure depends only on $M_0$
and $L_0$. This behavior explains that the correlation of $M_0$ with
the star radius shown in Fig.\,\ref{fig2} is better for $R_{1.4}$ than
for $R_{1.0}$, and also that the correlation of $R$ with $M_0+\beta
L_0$ is so strong; c) the contribution of the $K_0$ term in (\ref{KL})
is more important than the $Q_0$ term  for $x-1<1$, which explains the
correlation of $R$ with $K_0+\alpha L_0$; d)
d) the asymmetry parameter $\delta$ monotonically decreases from its maximum
value $\sim$ 0.95,
obtained at densities of the order of $\rho_0/2$ to $0.65<\delta<0.85$ at $2\rho_0$. 
If the term in $K_0$
is neglected in eq. (\ref{ML}), the pressure satisfies $P\propto M_0 +
\frac{18\delta^2}{(x-1)^2} L_0 =M_0+ \beta' L_0.$ Taking for $\beta'$
the upper and lower value of $\beta$ in Table\,\ref{tab2}, we get,
respectively, $x=1.45$ and $x=1.99$, above and just below the value $x=1.5$,
when the relation is exact. Therefore, it seems the relation is being
applied within the valid range of density. On the other hand, from
eq. (\ref{KL}) and neglecting the term in $Q_0$, the relation $P\propto
K_0+\frac{3\delta^2}{(x-1)}L_0= K_0+\alpha' L_0,$ is obtained.  We now
take for $\alpha'$ the upper and lower values of $\alpha$  from Table\,\ref{tab2},
 and we get, respectively, $x=1.49$ and $x=4.40$.
The last value is already out of the validity of the approximation, and even for $x=2$,
this approximation is not very good.  This might be the plausible reason
that only the low-mass neutron star radii are strongly correlated with
$K_0+\alpha L_0$.

\begin{figure}[tb!]
 \begin{center}
    \includegraphics[width=.4\textwidth,angle=0]{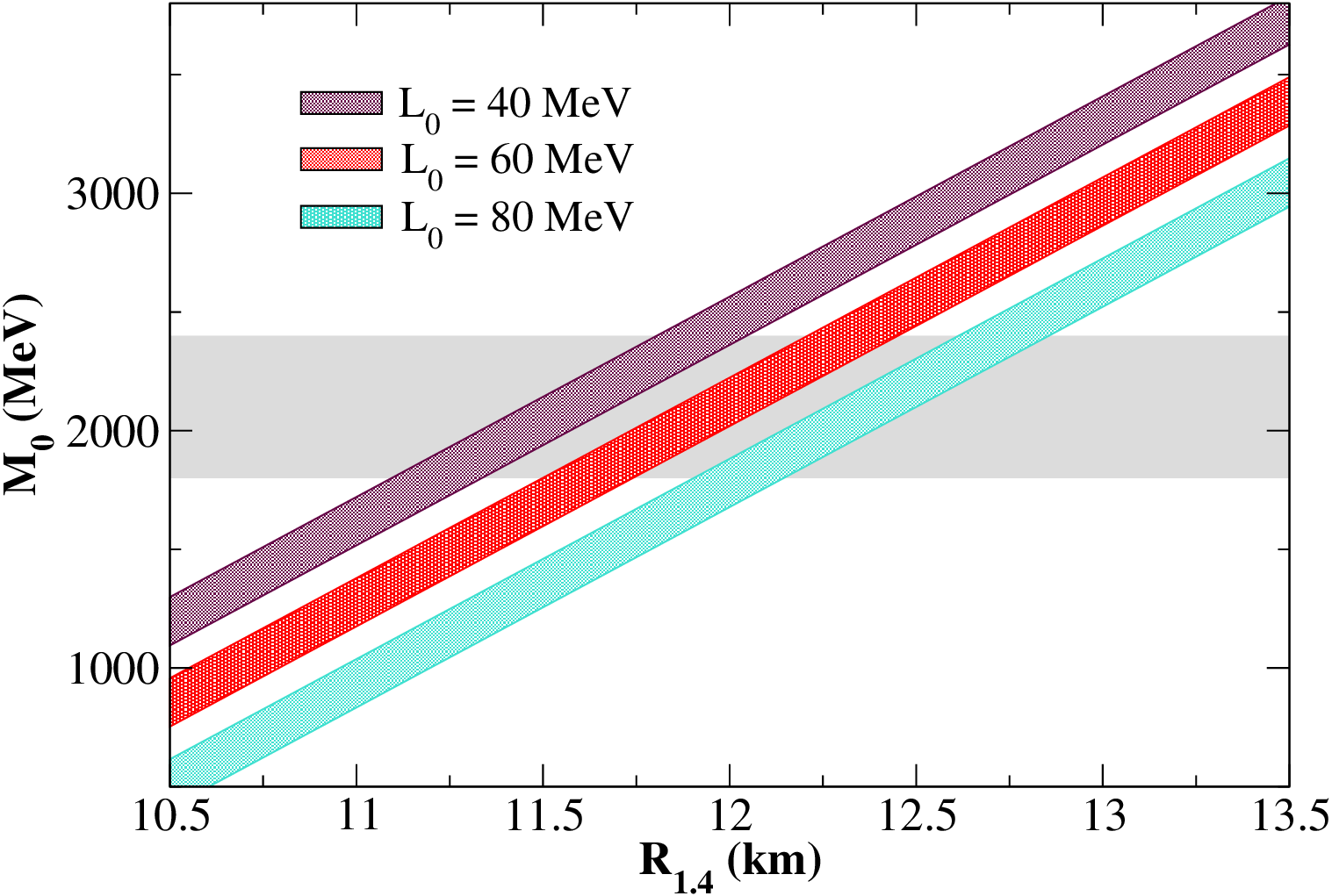}
     \end{center}
  \caption{(Color online) $M_0$ as a function of $R_{1.4}$ for $L_0=40$, $60$ and $80$ MeV, as obtained
  from the multiple linear regression. The gray shaded region indicates the
   constraint on $M_0$ derived in Ref. \cite{De15}.}
 \label{fig5}
\end{figure}

The knowledge of the slopes of nuclear matter incompressibility
and symmetry energy at  saturation density can  constrain the
neutron star radii as these radii  are strongly correlated with $M_0
+ \beta L_0$.  An overall variation in $L_0 =  20\text{-}80$ MeV is
obtained from the analysis of the giant dipole resonance of $^{208}$Pb
\cite{Trippa08, Roca-Maza13a}, the giant quadrupole resonance in
$^{208}$Pb \cite{Roca-maza13}, the pigmy dipole resonance in $^{68}$Ni
and $^{132}$Sn \cite{Carbone10}, and nuclear ground state properties,
using the standard Skyrme Hartree-Fock approach \cite{Chen11}.
A fit of the EoS for  asymmetric nuclear matter, or pure neutron matter
\cite{Chabanat98,Roca-Maza11}, or the binding energies of large number
of deformed nuclei \cite{Niksic08, Zhao10} within different mean field
models, constrains the value of $L_0$ in the range of  $40 - 70$ MeV. The
combined results of nuclear structure and heavy ion collisions lead to the
central value of $L_0=70$ MeV \cite{Tsang12}. We adopt $L_0=40\text{-}80$
MeV, which has a good overlap with these investigations.  The value of
$M_0 = 1800 \text{-} 2400$ MeV \cite{De15} at the saturation density seems
to be consistent with its value at $\rho = 0.1$ fm$^{-3}$, deduced from
the energies of the isoscalar giant monopole resonance in the $^{132}$Sn
and $^{208}$Pb nuclei \cite{Khan12,khan13}.  In Fig.\,\ref{fig5}, we plot
the incompressibility slope parameter $M_0$ as a function of $R_{1.4}$,
for different values of the symmetry energy slope $L_0=40$, $60$ and
$80$ MeV.  The gray shaded region corresponds to the constraint on
$M_0$ as obtained in Ref. \cite{De15}.  These values of $M_0$ and $L_0$
suggest that $R_{1.4}$ should be in the range of $11.09\mbox{-}12.86$
km, which is consistent with the results of the Ref. \cite{Steiner16}.
Let us remark that if we had  only taken the RMF models, the above
  correlations would have been slightly stronger, as expected, because all
  models in the study would have had a similar underlying framework, and
  larger radii would have been obtained for a 1.4 $M_\odot$ star, namely
  $R_{1.4}$ would have  come out  in the range of $11.82\mbox{-}13.25$
  km. Indeed Fig.\,\ref{fig1} shows that, on average,
   RMF models lead to larger radii than the other types of calculations.

In conclusion, we have studied the possible existing correlations between
neutron star radii at different masses and the nuclear coefficients of
the nuclear matter EoSs, calculated at the saturation density. The neutron
star radii are obtained using unified EoSs, fully for the Skyrme models,
and partially for inner-crust-core EoSs for the RMF models, except for
the two  microscopic EoSs. All EoSs are consistent with the existence of
$2M_\odot$ neutron stars.  The radii of the low-mass neutron stars are
better correlated with the symmetry energy coefficient $J_0$ and its slope
$L_0$. As the neutron star mass increases, the correlations of the radii
with the nuclear matter incompressibility $K_0$ and its slope
$M_0$ grow stronger.  Our investigation reveals that the neutron star
radii are better correlated with the linear combinations $K_0 + \alpha
L_0$ and $M_0 + \beta L_0$  than with the individual EoS parameters.
In particular, noticeable improvement is seen in the correlations
of the radii with these linear combinations, for $~1.4M_{\odot}$
neutron stars.  The correlations of the radii with $M_0 + \beta L_0$
are stronger, and almost independent of the neutron star mass, in the
range $0.6\text{-}1.8M_{\odot}$.  A plausible interpretation for the
existence of such correlations is traced back to the correlations between
the pressure and similar linear combinations of the EoS parameters in
the relevant density range. The values of $M_0$ and $L_0$, as 
currently deduced from finite nuclei data, constrain  $R_{1.4}$ in the
range $11.09\text{-}12.86$ km.

\vspace{0.5cm}
H.P. is supported by FCT under Project No. SFRH/BPD/95566/2013. Partial
support comes from ``NewCompStar'', COST Action MP1304. The work of M.F. has been partially supported by the NCN (Poland) Grant No. 2014/13/B/ST9/02621.

%\bibliography{plb_ref-1}

\begin{thebibliography}{73}%
\makeatletter
\providecommand \@ifxundefined [1]{%
 \@ifx{#1\undefined}
}%
\providecommand \@ifnum [1]{%
 \ifnum #1\expandafter \@firstoftwo
 \else \expandafter \@secondoftwo
 \fi
}%
\providecommand \@ifx [1]{%
 \ifx #1\expandafter \@firstoftwo
 \else \expandafter \@secondoftwo
 \fi
}%
\providecommand \natexlab [1]{#1}%
\providecommand \enquote  [1]{``#1''}%
\providecommand \bibnamefont  [1]{#1}%
\providecommand \bibfnamefont [1]{#1}%
\providecommand \citenamefont [1]{#1}%
\providecommand \href@noop [0]{\@secondoftwo}%
\providecommand \href [0]{\begingroup \@sanitize@url \@href}%
\providecommand \@href[1]{\@@startlink{#1}\@@href}%
\providecommand \@@href[1]{\endgroup#1\@@endlink}%
\providecommand \@sanitize@url [0]{\catcode `\\12\catcode `\$12\catcode
  `\&12\catcode `\#12\catcode `\^12\catcode `\_12\catcode `\%12\relax}%
\providecommand \@@startlink[1]{}%
\providecommand \@@endlink[0]{}%
\providecommand \url  [0]{\begingroup\@sanitize@url \@url }%
\providecommand \@url [1]{\endgroup\@href {#1}{\urlprefix }}%
\providecommand \urlprefix  [0]{URL }%
\providecommand \Eprint [0]{\href }%
\providecommand \doibase [0]{http://dx.doi.org/}%
\providecommand \selectlanguage [0]{\@gobble}%
\providecommand \bibinfo  [0]{\@secondoftwo}%
\providecommand \bibfield  [0]{\@secondoftwo}%
\providecommand \translation [1]{[#1]}%
\providecommand \BibitemOpen [0]{}%
\providecommand \bibitemStop [0]{}%
\providecommand \bibitemNoStop [0]{.\EOS\space}%
\providecommand \EOS [0]{\spacefactor3000\relax}%
\providecommand \BibitemShut  [1]{\csname bibitem#1\endcsname}%
\let\auto@bib@innerbib\@empty
%</preamble>
\bibitem [{\citenamefont {Glendenning}(1986)}]{Glendenning86}%
  \BibitemOpen
  \bibfield  {author} {\bibinfo {author} {\bibfnamefont {N.~K.}\ \bibnamefont
  {Glendenning}},\ }\href@noop {} {\bibfield  {journal} {\bibinfo  {journal}
  {Phys. Rev. Lett.}\ }\textbf {\bibinfo {volume} {57}},\ \bibinfo {pages}
  {1120} (\bibinfo {year} {1986})}\BibitemShut {NoStop}%
\bibitem [{\citenamefont {Horowitz}\ and\ \citenamefont
  {Piekarewicz}(2001)}]{Horowitz01}%
  \BibitemOpen
  \bibfield  {author} {\bibinfo {author} {\bibfnamefont {C.~J.}\ \bibnamefont
  {Horowitz}}\ and\ \bibinfo {author} {\bibfnamefont {J.}~\bibnamefont
  {Piekarewicz}},\ }\href@noop {} {\bibfield  {journal} {\bibinfo  {journal}
  {Phys. Rev. Lett.}\ }\textbf {\bibinfo {volume} {86}},\ \bibinfo {pages}
  {5647} (\bibinfo {year} {2001})}\BibitemShut {NoStop}%
\bibitem [{\citenamefont {Lattimer}\ and\ \citenamefont
  {Prakesh}(2007)}]{Lattimer07}%
  \BibitemOpen
  \bibfield  {author} {\bibinfo {author} {\bibfnamefont {J.~M.}\ \bibnamefont
  {Lattimer}}\ and\ \bibinfo {author} {\bibfnamefont {M.}~\bibnamefont
  {Prakesh}},\ }\href@noop {} {\bibfield  {journal} {\bibinfo  {journal} {Phys.
  Rep.}\ }\textbf {\bibinfo {volume} {442}},\ \bibinfo {pages} {109} (\bibinfo
  {year} {2007})}\BibitemShut {NoStop}%
\bibitem [{\citenamefont {Xu}\ \emph {et~al.}(2009)\citenamefont {Xu},
  \citenamefont {Chen}, \citenamefont {Li},\ and\ \citenamefont {Ma}}]{Xu09}%
  \BibitemOpen
  \bibfield  {author} {\bibinfo {author} {\bibfnamefont {J.}~\bibnamefont
  {Xu}}, \bibinfo {author} {\bibfnamefont {L.-W.}\ \bibnamefont {Chen}},
  \bibinfo {author} {\bibfnamefont {B.-A.}\ \bibnamefont {Li}}, \ and\ \bibinfo
  {author} {\bibfnamefont {H.-R.}\ \bibnamefont {Ma}},\ }\href@noop {}
  {\bibfield  {journal} {\bibinfo  {journal} {Astrophys. J.}\ }\textbf
  {\bibinfo {volume} {697}},\ \bibinfo {pages} {1549} (\bibinfo {year}
  {2009})}\BibitemShut {NoStop}%
\bibitem [{\citenamefont {Vida{\~n}a}\ \emph {et~al.}(2009)\citenamefont
  {Vida{\~n}a}, \citenamefont {Provid{\^e}ncia}, \citenamefont {Polls},\ and\
  \citenamefont {Rios}}]{vidana09}%
  \BibitemOpen
  \bibfield  {author} {\bibinfo {author} {\bibfnamefont {I.}~\bibnamefont
  {Vida{\~n}a}}, \bibinfo {author} {\bibfnamefont {C.}~\bibnamefont
  {Provid{\^e}ncia}}, \bibinfo {author} {\bibfnamefont {A.}~\bibnamefont
  {Polls}}, \ and\ \bibinfo {author} {\bibfnamefont {A.}~\bibnamefont {Rios}},\
  }\href@noop {} {\bibfield  {journal} {\bibinfo  {journal} {Phys. Rev. C}\
  }\textbf {\bibinfo {volume} {80}},\ \bibinfo {pages} {045806} (\bibinfo
  {year} {2009})}\BibitemShut {NoStop}%
\bibitem [{\citenamefont {{Ducoin}}\ \emph {et~al.}(2010)\citenamefont
  {{Ducoin}}, \citenamefont {{Margueron}},\ and\ \citenamefont
  {{Provid{\^e}ncia}}}]{Ducoin10}%
  \BibitemOpen
  \bibfield  {author} {\bibinfo {author} {\bibfnamefont {C.}~\bibnamefont
  {{Ducoin}}}, \bibinfo {author} {\bibfnamefont {J.}~\bibnamefont
  {{Margueron}}}, \ and\ \bibinfo {author} {\bibfnamefont {C.}~\bibnamefont
  {{Provid{\^e}ncia}}},\ }\href {\doibase 10.1209/0295-5075/91/32001}
  {\bibfield  {journal} {\bibinfo  {journal} {Europhysics Letters}\ }\textbf
  {\bibinfo {volume} {91}},\ \bibinfo {pages} {32001} (\bibinfo {year}
  {2010})}\BibitemShut {NoStop}%
\bibitem [{\citenamefont {Ducoin}\ \emph {et~al.}(2011)\citenamefont {Ducoin},
  \citenamefont {Margueron}, \citenamefont {Provid{\^e}ncia},\ and\
  \citenamefont {Vida{\~n}a}}]{Ducoin11}%
  \BibitemOpen
  \bibfield  {author} {\bibinfo {author} {\bibfnamefont {C.}~\bibnamefont
  {Ducoin}}, \bibinfo {author} {\bibfnamefont {J.}~\bibnamefont {Margueron}},
  \bibinfo {author} {\bibfnamefont {C.}~\bibnamefont {Provid{\^e}ncia}}, \ and\
  \bibinfo {author} {\bibfnamefont {I.}~\bibnamefont {Vida{\~n}a}},\
  }\href@noop {} {\bibfield  {journal} {\bibinfo  {journal} {Phys. Rev. C}\
  }\textbf {\bibinfo {volume} {83}},\ \bibinfo {pages} {045810} (\bibinfo
  {year} {2011})}\BibitemShut {NoStop}%
\bibitem [{\citenamefont {Cavagnoli}\ \emph {et~al.}(2011)\citenamefont
  {Cavagnoli}, \citenamefont {Menezes},\ and\ \citenamefont
  {Provid\^encia}}]{Cavagnoli11}%
  \BibitemOpen
  \bibfield  {author} {\bibinfo {author} {\bibfnamefont {R.}~\bibnamefont
  {Cavagnoli}}, \bibinfo {author} {\bibfnamefont {D.~P.}\ \bibnamefont
  {Menezes}}, \ and\ \bibinfo {author} {\bibfnamefont {C.}~\bibnamefont
  {Provid\^encia}},\ }\href@noop {} {\bibfield  {journal} {\bibinfo  {journal}
  {Phys. Rev. C}\ }\textbf {\bibinfo {volume} {84}},\ \bibinfo {pages} {065810}
  (\bibinfo {year} {2011})}\BibitemShut {NoStop}%
\bibitem [{\citenamefont {Gandolfi}\ \emph {et~al.}(2012)\citenamefont
  {Gandolfi}, \citenamefont {Carlson},\ and\ \citenamefont
  {Reddy}}]{Gandolfi12}%
  \BibitemOpen
  \bibfield  {author} {\bibinfo {author} {\bibfnamefont {S.}~\bibnamefont
  {Gandolfi}}, \bibinfo {author} {\bibfnamefont {J.}~\bibnamefont {Carlson}}, \
  and\ \bibinfo {author} {\bibfnamefont {S.}~\bibnamefont {Reddy}},\
  }\href@noop {} {\bibfield  {journal} {\bibinfo  {journal} {Phys. Rev. C}\
  }\textbf {\bibinfo {volume} {85}},\ \bibinfo {pages} {032801} (\bibinfo
  {year} {2012})}\BibitemShut {NoStop}%
\bibitem [{\citenamefont {Fattoyev}\ and\ \citenamefont
  {Piekarewicz}(2012)}]{Fattoyev12}%
  \BibitemOpen
  \bibfield  {author} {\bibinfo {author} {\bibfnamefont {F.~J.}\ \bibnamefont
  {Fattoyev}}\ and\ \bibinfo {author} {\bibfnamefont {J.}~\bibnamefont
  {Piekarewicz}},\ }\href@noop {} {\bibfield  {journal} {\bibinfo  {journal}
  {Phys. Rev. C}\ }\textbf {\bibinfo {volume} {86}},\ \bibinfo {pages} {015802}
  (\bibinfo {year} {2012})}\BibitemShut {NoStop}%
\bibitem [{\citenamefont {Newton}\ \emph {et~al.}(2013)\citenamefont {Newton},
  \citenamefont {Gearheart},\ and\ \citenamefont {Li}}]{Newton13}%
  \BibitemOpen
  \bibfield  {author} {\bibinfo {author} {\bibfnamefont {W.~G.}\ \bibnamefont
  {Newton}}, \bibinfo {author} {\bibfnamefont {M.}~\bibnamefont {Gearheart}}, \
  and\ \bibinfo {author} {\bibfnamefont {B.-A.}\ \bibnamefont {Li}},\
  }\href@noop {} {\bibfield  {journal} {\bibinfo  {journal} {Astrophys. J.
  Suppl.}\ }\textbf {\bibinfo {volume} {204}},\ \bibinfo {pages} {9} (\bibinfo
  {year} {2013})}\BibitemShut {NoStop}%
\bibitem [{\citenamefont {Fattoyev}\ \emph {et~al.}(2014)\citenamefont
  {Fattoyev}, \citenamefont {Newton},\ and\ \citenamefont {Li}}]{Fattoyev14}%
  \BibitemOpen
  \bibfield  {author} {\bibinfo {author} {\bibfnamefont {F.~J.}\ \bibnamefont
  {Fattoyev}}, \bibinfo {author} {\bibfnamefont {W.}~\bibnamefont {Newton}}, \
  and\ \bibinfo {author} {\bibfnamefont {B.-A.}\ \bibnamefont {Li}},\
  }\href@noop {} {\bibfield  {journal} {\bibinfo  {journal} {Phys. Rev. C}\
  }\textbf {\bibinfo {volume} {90}},\ \bibinfo {pages} {022801(R)} (\bibinfo
  {year} {2014})}\BibitemShut {NoStop}%
\bibitem [{\citenamefont {Sotani}\ \emph {et~al.}(2015)\citenamefont {Sotani},
  \citenamefont {Iida},\ and\ \citenamefont {Oyamatsu}}]{Sotani15}%
  \BibitemOpen
  \bibfield  {author} {\bibinfo {author} {\bibfnamefont {H.}~\bibnamefont
  {Sotani}}, \bibinfo {author} {\bibfnamefont {K.}~\bibnamefont {Iida}}, \ and\
  \bibinfo {author} {\bibfnamefont {K.}~\bibnamefont {Oyamatsu}},\ }\href@noop
  {} {\bibfield  {journal} {\bibinfo  {journal} {Phys. Rev. C}\ }\textbf
  {\bibinfo {volume} {91}},\ \bibinfo {pages} {015805} (\bibinfo {year}
  {2015})}\BibitemShut {NoStop}%
\bibitem [{\citenamefont {Fortin}\ \emph {et~al.}(2016)\citenamefont {Fortin},
  \citenamefont {Provid\^encia}, \citenamefont {Raduta}, \citenamefont
  {Gulminelli}, \citenamefont {Zdunik}, \citenamefont {Haensel},\ and\
  \citenamefont {Bejger}}]{Fortin16}%
  \BibitemOpen
  \bibfield  {author} {\bibinfo {author} {\bibfnamefont {M.}~\bibnamefont
  {Fortin}}, \bibinfo {author} {\bibfnamefont {C.}~\bibnamefont
  {Provid\^encia}}, \bibinfo {author} {\bibfnamefont {A.~R.}\ \bibnamefont
  {Raduta}}, \bibinfo {author} {\bibfnamefont {F.}~\bibnamefont {Gulminelli}},
  \bibinfo {author} {\bibfnamefont {J.~L.}\ \bibnamefont {Zdunik}}, \bibinfo
  {author} {\bibfnamefont {P.}~\bibnamefont {Haensel}}, \ and\ \bibinfo
  {author} {\bibfnamefont {M.}~\bibnamefont {Bejger}},\ }\href {\doibase
  10.1103/PhysRevC.94.035804} {\bibfield  {journal} {\bibinfo  {journal} {Phys.
  Rev. C}\ }\textbf {\bibinfo {volume} {94}},\ \bibinfo {pages} {035804}
  (\bibinfo {year} {2016})}\BibitemShut {NoStop}%
\bibitem [{\citenamefont {Dutra}\ \emph {et~al.}(2016)\citenamefont {Dutra},
  \citenamefont {Louren\ifmmode~\mbox{\c{c}}\else \c{c}\fi{}o},\ and\
  \citenamefont {Menezes}}]{Dutra16}%
  \BibitemOpen
  \bibfield  {author} {\bibinfo {author} {\bibfnamefont {M.}~\bibnamefont
  {Dutra}}, \bibinfo {author} {\bibfnamefont {O.}~\bibnamefont
  {Louren\ifmmode~\mbox{\c{c}}\else \c{c}\fi{}o}}, \ and\ \bibinfo {author}
  {\bibfnamefont {D.~P.}\ \bibnamefont {Menezes}},\ }\href@noop {} {\bibfield
  {journal} {\bibinfo  {journal} {Phys. Rev. C}\ }\textbf {\bibinfo {volume}
  {93}},\ \bibinfo {pages} {025806} (\bibinfo {year} {2016})}\BibitemShut
  {NoStop}%
\bibitem [{\citenamefont {Taranto}\ \emph {et~al.}(2013)\citenamefont
  {Taranto}, \citenamefont {Baldo},\ and\ \citenamefont {Burgio}}]{Taranto13}%
  \BibitemOpen
  \bibfield  {author} {\bibinfo {author} {\bibfnamefont {G.}~\bibnamefont
  {Taranto}}, \bibinfo {author} {\bibfnamefont {M.}~\bibnamefont {Baldo}}, \
  and\ \bibinfo {author} {\bibfnamefont {G.~F.}\ \bibnamefont {Burgio}},\
  }\href {\doibase 10.1103/PhysRevC.87.045803} {\bibfield  {journal} {\bibinfo
  {journal} {Phys. Rev.}\ }\textbf {\bibinfo {volume} {C87}},\ \bibinfo {pages}
  {045803} (\bibinfo {year} {2013})}\BibitemShut {NoStop}%
\bibitem [{\citenamefont {Akmal}\ \emph {et~al.}(1998)\citenamefont {Akmal},
  \citenamefont {Pandharipande},\ and\ \citenamefont {Ravenhall}}]{APR}%
  \BibitemOpen
  \bibfield  {author} {\bibinfo {author} {\bibfnamefont {A.}~\bibnamefont
  {Akmal}}, \bibinfo {author} {\bibfnamefont {V.~R.}\ \bibnamefont
  {Pandharipande}}, \ and\ \bibinfo {author} {\bibfnamefont {D.~G.}\
  \bibnamefont {Ravenhall}},\ }\href {\doibase 10.1103/PhysRevC.58.1804}
  {\bibfield  {journal} {\bibinfo  {journal} {Phys. Rev. C}\ }\textbf {\bibinfo
  {volume} {58}},\ \bibinfo {pages} {1804} (\bibinfo {year}
  {1998})}\BibitemShut {NoStop}%
\bibitem [{\citenamefont {Alam}\ \emph {et~al.}(2014)\citenamefont {Alam},
  \citenamefont {Agrawal}, \citenamefont {De}, \citenamefont {Samaddar},\ and\
  \citenamefont {Col\`o}}]{Alam15}%
  \BibitemOpen
  \bibfield  {author} {\bibinfo {author} {\bibfnamefont {N.}~\bibnamefont
  {Alam}}, \bibinfo {author} {\bibfnamefont {B.~K.}\ \bibnamefont {Agrawal}},
  \bibinfo {author} {\bibfnamefont {J.~N.}\ \bibnamefont {De}}, \bibinfo
  {author} {\bibfnamefont {S.~K.}\ \bibnamefont {Samaddar}}, \ and\ \bibinfo
  {author} {\bibfnamefont {G.}~\bibnamefont {Col\`o}},\ }\href@noop {}
  {\bibfield  {journal} {\bibinfo  {journal} {Phys. Rev. C}\ }\textbf {\bibinfo
  {volume} {90}},\ \bibinfo {pages} {054317} (\bibinfo {year}
  {2014})}\BibitemShut {NoStop}%
\bibitem [{\citenamefont {Dhiman}\ \emph {et~al.}(2007)\citenamefont {Dhiman},
  \citenamefont {Kumar},\ and\ \citenamefont {Agrawal}}]{Dhiman07}%
  \BibitemOpen
  \bibfield  {author} {\bibinfo {author} {\bibfnamefont {S.~K.}\ \bibnamefont
  {Dhiman}}, \bibinfo {author} {\bibfnamefont {R.}~\bibnamefont {Kumar}}, \
  and\ \bibinfo {author} {\bibfnamefont {B.~K.}\ \bibnamefont {Agrawal}},\
  }\href@noop {} {\bibfield  {journal} {\bibinfo  {journal} {Phys. Rev. C}\
  }\textbf {\bibinfo {volume} {76}},\ \bibinfo {pages} {045801} (\bibinfo
  {year} {2007})}\BibitemShut {NoStop}%
\bibitem [{\citenamefont {Agrawal}(2010)}]{Agrawal10}%
  \BibitemOpen
  \bibfield  {author} {\bibinfo {author} {\bibfnamefont {B.~K.}\ \bibnamefont
  {Agrawal}},\ }\href@noop {} {\bibfield  {journal} {\bibinfo  {journal} {Phys.
  Rev. C}\ }\textbf {\bibinfo {volume} {81}},\ \bibinfo {pages} {034323}
  (\bibinfo {year} {2010})}\BibitemShut {NoStop}%
\bibitem [{\citenamefont {Chen}\ and\ \citenamefont
  {Piekarewicz}(2014)}]{Chen14}%
  \BibitemOpen
  \bibfield  {author} {\bibinfo {author} {\bibfnamefont {W.-C.}\ \bibnamefont
  {Chen}}\ and\ \bibinfo {author} {\bibfnamefont {J.}~\bibnamefont
  {Piekarewicz}},\ }\href@noop {} {\bibfield  {journal} {\bibinfo  {journal}
  {Phys. Rev. C}\ }\textbf {\bibinfo {volume} {90}},\ \bibinfo {pages} {044305}
  (\bibinfo {year} {2014})}\BibitemShut {NoStop}%
\bibitem [{\citenamefont {Glendenning}\ and\ \citenamefont
  {Moszkowski}(1991)}]{Glendenning91}%
  \BibitemOpen
  \bibfield  {author} {\bibinfo {author} {\bibfnamefont {N.~K.}\ \bibnamefont
  {Glendenning}}\ and\ \bibinfo {author} {\bibfnamefont {S.~A.}\ \bibnamefont
  {Moszkowski}},\ }\href@noop {} {\bibfield  {journal} {\bibinfo  {journal}
  {Phys. Rev. Lett.}\ }\textbf {\bibinfo {volume} {67}},\ \bibinfo {pages}
  {2414} (\bibinfo {year} {1991})}\BibitemShut {NoStop}%
\bibitem [{\citenamefont {Lalazissis}\ \emph {et~al.}(1997)\citenamefont
  {Lalazissis}, \citenamefont {K\"onig},\ and\ \citenamefont
  {Ring}}]{Lalazissis97}%
  \BibitemOpen
  \bibfield  {author} {\bibinfo {author} {\bibfnamefont {G.~A.}\ \bibnamefont
  {Lalazissis}}, \bibinfo {author} {\bibfnamefont {J.}~\bibnamefont {K\"onig}},
  \ and\ \bibinfo {author} {\bibfnamefont {P.}~\bibnamefont {Ring}},\
  }\href@noop {} {\bibfield  {journal} {\bibinfo  {journal} {Phys. Rev. C}\
  }\textbf {\bibinfo {volume} {55}},\ \bibinfo {pages} {540} (\bibinfo {year}
  {1997})}\BibitemShut {NoStop}%
\bibitem [{\citenamefont {Pais}\ and\ \citenamefont
  {Provid{\^e}ncia}(2016)}]{Pais16}%
  \BibitemOpen
  \bibfield  {author} {\bibinfo {author} {\bibfnamefont {H.}~\bibnamefont
  {Pais}}\ and\ \bibinfo {author} {\bibfnamefont {C.}~\bibnamefont
  {Provid{\^e}ncia}},\ }\href@noop {} {\bibfield  {journal} {\bibinfo
  {journal} {Phys. Rev. C}\ }\textbf {\bibinfo {volume} {94}},\ \bibinfo
  {pages} {015808} (\bibinfo {year} {2016})}\BibitemShut {NoStop}%
\bibitem [{\citenamefont {Carriere}\ \emph {et~al.}(2003)\citenamefont
  {Carriere}, \citenamefont {Horowitz},\ and\ \citenamefont
  {Piekarewicz}}]{Carriere03}%
  \BibitemOpen
  \bibfield  {author} {\bibinfo {author} {\bibfnamefont {J.}~\bibnamefont
  {Carriere}}, \bibinfo {author} {\bibfnamefont {C.}~\bibnamefont {Horowitz}},
  \ and\ \bibinfo {author} {\bibfnamefont {J.}~\bibnamefont {Piekarewicz}},\
  }\href@noop {} {\bibfield  {journal} {\bibinfo  {journal} {Astrophys. J.}\
  }\textbf {\bibinfo {volume} {593}},\ \bibinfo {pages} {463} (\bibinfo {year}
  {2003})}\BibitemShut {NoStop}%
\bibitem [{\citenamefont {Sugahara}\ and\ \citenamefont
  {Toki}(1994)}]{Sugahara94}%
  \BibitemOpen
  \bibfield  {author} {\bibinfo {author} {\bibfnamefont {Y.}~\bibnamefont
  {Sugahara}}\ and\ \bibinfo {author} {\bibfnamefont {H.}~\bibnamefont
  {Toki}},\ }\href@noop {} {\bibfield  {journal} {\bibinfo  {journal} {Nucl.
  Phys. A}\ }\textbf {\bibinfo {volume} {579}},\ \bibinfo {pages} {557}
  (\bibinfo {year} {1994})}\BibitemShut {NoStop}%
\bibitem [{\citenamefont {Provid{\^e}ncia}\ and\ \citenamefont
  {Rabhi}(2013)}]{Providencia13}%
  \BibitemOpen
  \bibfield  {author} {\bibinfo {author} {\bibfnamefont {C.}~\bibnamefont
  {Provid{\^e}ncia}}\ and\ \bibinfo {author} {\bibfnamefont {A.}~\bibnamefont
  {Rabhi}},\ }\href@noop {} {\bibfield  {journal} {\bibinfo  {journal} {Phys.
  Rev. C}\ }\textbf {\bibinfo {volume} {87}},\ \bibinfo {pages} {055801}
  (\bibinfo {year} {2013})}\BibitemShut {NoStop}%
\bibitem [{\citenamefont {Typel}\ \emph {et~al.}(2010)\citenamefont {Typel},
  \citenamefont {Ropke}, \citenamefont {Klahn}, \citenamefont {Blaschke},\ and\
  \citenamefont {Wolter}}]{Typel10}%
  \BibitemOpen
  \bibfield  {author} {\bibinfo {author} {\bibfnamefont {S.}~\bibnamefont
  {Typel}}, \bibinfo {author} {\bibfnamefont {G.}~\bibnamefont {Ropke}},
  \bibinfo {author} {\bibfnamefont {T.}~\bibnamefont {Klahn}}, \bibinfo
  {author} {\bibfnamefont {D.}~\bibnamefont {Blaschke}}, \ and\ \bibinfo
  {author} {\bibfnamefont {H.~H.}\ \bibnamefont {Wolter}},\ }\href@noop {}
  {\bibfield  {journal} {\bibinfo  {journal} {Phys. Rev. C}\ }\textbf {\bibinfo
  {volume} {81}},\ \bibinfo {pages} {015803} (\bibinfo {year}
  {2010})}\BibitemShut {NoStop}%
\bibitem [{\citenamefont {Gaitanos}\ \emph {et~al.}(2004)\citenamefont
  {Gaitanos}, \citenamefont {Di~Toro}, \citenamefont {Typel}, \citenamefont
  {Baran}, \citenamefont {Fuchs}, \citenamefont {Greco},\ and\ \citenamefont
  {Wolter}}]{Gaitanos04}%
  \BibitemOpen
  \bibfield  {author} {\bibinfo {author} {\bibfnamefont {T.}~\bibnamefont
  {Gaitanos}}, \bibinfo {author} {\bibfnamefont {M.}~\bibnamefont {Di~Toro}},
  \bibinfo {author} {\bibfnamefont {S.}~\bibnamefont {Typel}}, \bibinfo
  {author} {\bibfnamefont {V.}~\bibnamefont {Baran}}, \bibinfo {author}
  {\bibfnamefont {C.}~\bibnamefont {Fuchs}}, \bibinfo {author} {\bibfnamefont
  {V.}~\bibnamefont {Greco}}, \ and\ \bibinfo {author} {\bibfnamefont {H.~H.}\
  \bibnamefont {Wolter}},\ }\href@noop {} {\bibfield  {journal} {\bibinfo
  {journal} {Nucl. Phys.}\ }\textbf {\bibinfo {volume} {A732}},\ \bibinfo
  {pages} {24} (\bibinfo {year} {2004})}\BibitemShut {NoStop}%
\bibitem [{\citenamefont {Nik{\v s}i{\'c}}\ \emph {et~al.}(2002)\citenamefont
  {Nik{\v s}i{\'c}}, \citenamefont {Vretenar}, \citenamefont {Finelli},\ and\
  \citenamefont {Ring}}]{Niksic02}%
  \BibitemOpen
  \bibfield  {author} {\bibinfo {author} {\bibfnamefont {T.}~\bibnamefont
  {Nik{\v s}i{\'c}}}, \bibinfo {author} {\bibfnamefont {D.}~\bibnamefont
  {Vretenar}}, \bibinfo {author} {\bibfnamefont {P.}~\bibnamefont {Finelli}}, \
  and\ \bibinfo {author} {\bibfnamefont {P.}~\bibnamefont {Ring}},\ }\href@noop
  {} {\bibfield  {journal} {\bibinfo  {journal} {Phys. Rev. C}\ }\textbf
  {\bibinfo {volume} {66}},\ \bibinfo {pages} {024306} (\bibinfo {year}
  {2002})}\BibitemShut {NoStop}%
\bibitem [{\citenamefont {Lalazissis}\ \emph {et~al.}(2005)\citenamefont
  {Lalazissis}, \citenamefont {Nik{\v s}i{\'c}}, \citenamefont {Vretenar},\
  and\ \citenamefont {Ring}}]{Lalazissis05}%
  \BibitemOpen
  \bibfield  {author} {\bibinfo {author} {\bibfnamefont {G.~A.}\ \bibnamefont
  {Lalazissis}}, \bibinfo {author} {\bibfnamefont {T.}~\bibnamefont {Nik{\v
  s}i{\'c}}}, \bibinfo {author} {\bibfnamefont {D.}~\bibnamefont {Vretenar}}, \
  and\ \bibinfo {author} {\bibfnamefont {P.}~\bibnamefont {Ring}},\ }\href@noop
  {} {\bibfield  {journal} {\bibinfo  {journal} {Phys. Rev. C}\ }\textbf
  {\bibinfo {volume} {71}},\ \bibinfo {pages} {024312} (\bibinfo {year}
  {2005})}\BibitemShut {NoStop}%
\bibitem [{\citenamefont {Typel}\ and\ \citenamefont {Wolter}(1999)}]{Typel99}%
  \BibitemOpen
  \bibfield  {author} {\bibinfo {author} {\bibfnamefont {S.}~\bibnamefont
  {Typel}}\ and\ \bibinfo {author} {\bibfnamefont {H.~H.}\ \bibnamefont
  {Wolter}},\ }\href@noop {} {\bibfield  {journal} {\bibinfo  {journal} {Nucl.
  Phys. A}\ }\textbf {\bibinfo {volume} {656}},\ \bibinfo {pages} {331}
  (\bibinfo {year} {1999})}\BibitemShut {NoStop}%
\bibitem [{\citenamefont {Kohler}(1976)}]{Kohler76}%
  \BibitemOpen
  \bibfield  {author} {\bibinfo {author} {\bibfnamefont {H.}~\bibnamefont
  {Kohler}},\ }\href@noop {} {\bibfield  {journal} {\bibinfo  {journal}
  {Nuclear Physics A}\ }\textbf {\bibinfo {volume} {258}},\ \bibinfo {pages}
  {301} (\bibinfo {year} {1976})}\BibitemShut {NoStop}%
\bibitem [{\citenamefont {Reinhard}\ and\ \citenamefont
  {Flocard}(1995)}]{Reinhard95}%
  \BibitemOpen
  \bibfield  {author} {\bibinfo {author} {\bibfnamefont {P.-G.}\ \bibnamefont
  {Reinhard}}\ and\ \bibinfo {author} {\bibfnamefont {H.}~\bibnamefont
  {Flocard}},\ }\href@noop {} {\bibfield  {journal} {\bibinfo  {journal} {Nucl.
  Phys. A}\ }\textbf {\bibinfo {volume} {584}},\ \bibinfo {pages} {467}
  (\bibinfo {year} {1995})}\BibitemShut {NoStop}%
\bibitem [{\citenamefont {{Nazarewicz}}\ \emph {et~al.}(1996)\citenamefont
  {{Nazarewicz}}, \citenamefont {{Dobaczewski}}, \citenamefont {{Werner}},
  \citenamefont {{Maruhn}}, \citenamefont {{Reinhard}}, \citenamefont {{Rutz}},
  \citenamefont {{Chinn}}, \citenamefont {{Umar}},\ and\ \citenamefont
  {{Strayer}}}]{Nazarewicz96}%
  \BibitemOpen
  \bibfield  {author} {\bibinfo {author} {\bibfnamefont {W.}~\bibnamefont
  {{Nazarewicz}}}, \bibinfo {author} {\bibfnamefont {J.}~\bibnamefont
  {{Dobaczewski}}}, \bibinfo {author} {\bibfnamefont {T.~R.}\ \bibnamefont
  {{Werner}}}, \bibinfo {author} {\bibfnamefont {J.~A.}\ \bibnamefont
  {{Maruhn}}}, \bibinfo {author} {\bibfnamefont {P.-G.}\ \bibnamefont
  {{Reinhard}}}, \bibinfo {author} {\bibfnamefont {K.}~\bibnamefont {{Rutz}}},
  \bibinfo {author} {\bibfnamefont {C.~R.}\ \bibnamefont {{Chinn}}}, \bibinfo
  {author} {\bibfnamefont {A.~S.}\ \bibnamefont {{Umar}}}, \ and\ \bibinfo
  {author} {\bibfnamefont {M.~R.}\ \bibnamefont {{Strayer}}},\ }\href@noop {}
  {\bibfield  {journal} {\bibinfo  {journal} {Phys. Rev. C}\ }\textbf {\bibinfo
  {volume} {53}},\ \bibinfo {pages} {740} (\bibinfo {year} {1996})}\BibitemShut
  {NoStop}%
\bibitem [{\citenamefont {Chabanat}(1995)}]{ChabanatPhd}%
  \BibitemOpen
  \bibfield  {author} {\bibinfo {author} {\bibfnamefont {E.}~\bibnamefont
  {Chabanat}},\ }\href@noop {} {Ph.D. thesis},\ \bibinfo  {school} {University
  Claude Bernard Lyon-1, Lyon, France} (\bibinfo {year} {1995})\BibitemShut
  {NoStop}%
\bibitem [{\citenamefont {Chabanat}\ \emph {et~al.}(1997)\citenamefont
  {Chabanat}, \citenamefont {Bonche}, \citenamefont {Haensel}, \citenamefont
  {Meyer},\ and\ \citenamefont {Schaeffer}}]{Chabanat97}%
  \BibitemOpen
  \bibfield  {author} {\bibinfo {author} {\bibfnamefont {E.}~\bibnamefont
  {Chabanat}}, \bibinfo {author} {\bibfnamefont {P.}~\bibnamefont {Bonche}},
  \bibinfo {author} {\bibfnamefont {P.}~\bibnamefont {Haensel}}, \bibinfo
  {author} {\bibfnamefont {J.}~\bibnamefont {Meyer}}, \ and\ \bibinfo {author}
  {\bibfnamefont {R.}~\bibnamefont {Schaeffer}},\ }\href@noop {} {\bibfield
  {journal} {\bibinfo  {journal} {Nucl. Phys. A}\ }\textbf {\bibinfo {volume}
  {627}},\ \bibinfo {pages} {710} (\bibinfo {year} {1997})}\BibitemShut
  {NoStop}%
\bibitem [{\citenamefont {Chabanat}\ \emph {et~al.}(1998)\citenamefont
  {Chabanat}, \citenamefont {Bonche}, \citenamefont {Haensel}, \citenamefont
  {Meyer},\ and\ \citenamefont {Schaeffer}}]{Chabanat98}%
  \BibitemOpen
  \bibfield  {author} {\bibinfo {author} {\bibfnamefont {E.}~\bibnamefont
  {Chabanat}}, \bibinfo {author} {\bibfnamefont {P.}~\bibnamefont {Bonche}},
  \bibinfo {author} {\bibfnamefont {P.}~\bibnamefont {Haensel}}, \bibinfo
  {author} {\bibfnamefont {J.}~\bibnamefont {Meyer}}, \ and\ \bibinfo {author}
  {\bibfnamefont {R.}~\bibnamefont {Schaeffer}},\ }\href@noop {} {\bibfield
  {journal} {\bibinfo  {journal} {Nucl. Phys. A}\ }\textbf {\bibinfo {volume}
  {635}},\ \bibinfo {pages} {231} (\bibinfo {year} {1998})}\BibitemShut
  {NoStop}%
\bibitem [{\citenamefont {Bennour}\ \emph {et~al.}(1989)\citenamefont
  {Bennour}, \citenamefont {Heenen}, \citenamefont {Bonche}, \citenamefont
  {Dobaczewski},\ and\ \citenamefont {Flocard}}]{Bennour89}%
  \BibitemOpen
  \bibfield  {author} {\bibinfo {author} {\bibfnamefont {L.}~\bibnamefont
  {Bennour}}, \bibinfo {author} {\bibfnamefont {P.-H.}\ \bibnamefont {Heenen}},
  \bibinfo {author} {\bibfnamefont {P.}~\bibnamefont {Bonche}}, \bibinfo
  {author} {\bibfnamefont {J.}~\bibnamefont {Dobaczewski}}, \ and\ \bibinfo
  {author} {\bibfnamefont {H.}~\bibnamefont {Flocard}},\ }\href@noop {}
  {\bibfield  {journal} {\bibinfo  {journal} {Phys. Rev. C}\ }\textbf {\bibinfo
  {volume} {40}},\ \bibinfo {pages} {2834} (\bibinfo {year}
  {1989})}\BibitemShut {NoStop}%
\bibitem [{\citenamefont {Reinhard}(1999)}]{Reinhard99}%
  \BibitemOpen
  \bibfield  {author} {\bibinfo {author} {\bibfnamefont {P.~G.}\ \bibnamefont
  {Reinhard}},\ }\href@noop {} {\bibfield  {journal} {\bibinfo  {journal}
  {Nucl. Phys.}\ }\textbf {\bibinfo {volume} {A649}},\ \bibinfo {pages} {305c}
  (\bibinfo {year} {1999})}\BibitemShut {NoStop}%
\bibitem [{\citenamefont {Agrawal}\ \emph {et~al.}(2005)\citenamefont
  {Agrawal}, \citenamefont {Shlomo},\ and\ \citenamefont {Au}}]{Agrawal05}%
  \BibitemOpen
  \bibfield  {author} {\bibinfo {author} {\bibfnamefont {B.~K.}\ \bibnamefont
  {Agrawal}}, \bibinfo {author} {\bibfnamefont {S.}~\bibnamefont {Shlomo}}, \
  and\ \bibinfo {author} {\bibfnamefont {V.~K.}\ \bibnamefont {Au}},\
  }\href@noop {} {\bibfield  {journal} {\bibinfo  {journal} {Phys. Rev. C}\
  }\textbf {\bibinfo {volume} {72}},\ \bibinfo {pages} {014310} (\bibinfo
  {year} {2005})}\BibitemShut {NoStop}%
\bibitem [{\citenamefont {Agrawal}\ \emph {et~al.}(2003)\citenamefont
  {Agrawal}, \citenamefont {Shlomo},\ and\ \citenamefont {{Kim
  Au}}}]{Agrawal03}%
  \BibitemOpen
  \bibfield  {author} {\bibinfo {author} {\bibfnamefont {B.~K.}\ \bibnamefont
  {Agrawal}}, \bibinfo {author} {\bibfnamefont {S.}~\bibnamefont {Shlomo}}, \
  and\ \bibinfo {author} {\bibfnamefont {V.}~\bibnamefont {{Kim Au}}},\
  }\href@noop {} {\bibfield  {journal} {\bibinfo  {journal} {Phys. Rev. C}\
  }\textbf {\bibinfo {volume} {68}},\ \bibinfo {pages} {031304} (\bibinfo
  {year} {2003})}\BibitemShut {NoStop}%
\bibitem [{\citenamefont {Friedrich}\ and\ \citenamefont
  {Reinhard}(1986)}]{Friedrich86}%
  \BibitemOpen
  \bibfield  {author} {\bibinfo {author} {\bibfnamefont {J.}~\bibnamefont
  {Friedrich}}\ and\ \bibinfo {author} {\bibfnamefont {P.-G.}\ \bibnamefont
  {Reinhard}},\ }\href@noop {} {\bibfield  {journal} {\bibinfo  {journal}
  {Phys. Rev. C}\ }\textbf {\bibinfo {volume} {33}},\ \bibinfo {pages} {335}
  (\bibinfo {year} {1986})}\BibitemShut {NoStop}%
\bibitem [{\citenamefont {Goriely}\ \emph {et~al.}(2010)\citenamefont
  {Goriely}, \citenamefont {Chamel},\ and\ \citenamefont
  {Pearson}}]{Goriely10}%
  \BibitemOpen
  \bibfield  {author} {\bibinfo {author} {\bibfnamefont {S.}~\bibnamefont
  {Goriely}}, \bibinfo {author} {\bibfnamefont {N.}~\bibnamefont {Chamel}}, \
  and\ \bibinfo {author} {\bibfnamefont {J.~M.}\ \bibnamefont {Pearson}},\
  }\href@noop {} {\bibfield  {journal} {\bibinfo  {journal} {Phys. Rev. C}\
  }\textbf {\bibinfo {volume} {82}},\ \bibinfo {pages} {035804} (\bibinfo
  {year} {2010})}\BibitemShut {NoStop}%
\bibitem [{\citenamefont {Goriely}\ \emph {et~al.}(2013)\citenamefont
  {Goriely}, \citenamefont {Chamel},\ and\ \citenamefont
  {Pearson}}]{Goriely13}%
  \BibitemOpen
  \bibfield  {author} {\bibinfo {author} {\bibfnamefont {S.}~\bibnamefont
  {Goriely}}, \bibinfo {author} {\bibfnamefont {N.}~\bibnamefont {Chamel}}, \
  and\ \bibinfo {author} {\bibfnamefont {J.~M.}\ \bibnamefont {Pearson}},\
  }\href@noop {} {\bibfield  {journal} {\bibinfo  {journal} {Phys. Rev. C}\
  }\textbf {\bibinfo {volume} {88}},\ \bibinfo {pages} {024308} (\bibinfo
  {year} {2013})}\BibitemShut {NoStop}%
\bibitem [{\citenamefont {{Davesne}}\ \emph {et~al.}(2016)\citenamefont
  {{Davesne}}, \citenamefont {{Pastore}},\ and\ \citenamefont
  {{Navarro}}}]{Davesne16}%
  \BibitemOpen
  \bibfield  {author} {\bibinfo {author} {\bibfnamefont {D.}~\bibnamefont
  {{Davesne}}}, \bibinfo {author} {\bibfnamefont {A.}~\bibnamefont
  {{Pastore}}}, \ and\ \bibinfo {author} {\bibfnamefont {J.}~\bibnamefont
  {{Navarro}}},\ }\href {\doibase 10.1051/0004-6361/201526720} {\bibfield
  {journal} {\bibinfo  {journal} {Astron. Astrophys.}\ }\textbf {\bibinfo
  {volume} {585}},\ \bibinfo {eid} {A83} (\bibinfo {year} {2016})},\ \Eprint
  {http://arxiv.org/abs/1509.05744} {arXiv:1509.05744 [nucl-th]} \BibitemShut
  {NoStop}%
\bibitem [{\citenamefont {Steiner}\ \emph {et~al.}(2005)\citenamefont
  {Steiner}, \citenamefont {Prakash}, \citenamefont {Lattimer},\ and\
  \citenamefont {Ellis}}]{Steiner05}%
  \BibitemOpen
  \bibfield  {author} {\bibinfo {author} {\bibfnamefont {A.~W.}\ \bibnamefont
  {Steiner}}, \bibinfo {author} {\bibfnamefont {M.}~\bibnamefont {Prakash}},
  \bibinfo {author} {\bibfnamefont {J.~M.}\ \bibnamefont {Lattimer}}, \ and\
  \bibinfo {author} {\bibfnamefont {P.~J.}\ \bibnamefont {Ellis}},\ }\href@noop
  {} {\bibfield  {journal} {\bibinfo  {journal} {Phys. Rep.}\ }\textbf
  {\bibinfo {volume} {411}},\ \bibinfo {pages} {325} (\bibinfo {year}
  {2005})}\BibitemShut {NoStop}%
\bibitem [{\citenamefont {Baym}\ \emph {et~al.}(1971)\citenamefont {Baym},
  \citenamefont {Pethick},\ and\ \citenamefont {Sutherland}}]{Baym71}%
  \BibitemOpen
  \bibfield  {author} {\bibinfo {author} {\bibfnamefont {G.}~\bibnamefont
  {Baym}}, \bibinfo {author} {\bibfnamefont {C.}~\bibnamefont {Pethick}}, \
  and\ \bibinfo {author} {\bibfnamefont {P.}~\bibnamefont {Sutherland}},\
  }\href@noop {} {\bibfield  {journal} {\bibinfo  {journal} {Astrophys. J.}\
  }\textbf {\bibinfo {volume} {170}},\ \bibinfo {pages} {299} (\bibinfo {year}
  {1971})}\BibitemShut {NoStop}%
\bibitem [{\citenamefont {Grill}\ \emph {et~al.}(2014)\citenamefont {Grill},
  \citenamefont {Pais}, \citenamefont {Provid{\^e}ncia}, \citenamefont
  {Vida{\~n}a},\ and\ \citenamefont {Avancini}}]{Grill14}%
  \BibitemOpen
  \bibfield  {author} {\bibinfo {author} {\bibfnamefont {F.}~\bibnamefont
  {Grill}}, \bibinfo {author} {\bibfnamefont {H.}~\bibnamefont {Pais}},
  \bibinfo {author} {\bibfnamefont {C.}~\bibnamefont {Provid{\^e}ncia}},
  \bibinfo {author} {\bibfnamefont {I.}~\bibnamefont {Vida{\~n}a}}, \ and\
  \bibinfo {author} {\bibfnamefont {S.~S.}\ \bibnamefont {Avancini}},\
  }\href@noop {} {\bibfield  {journal} {\bibinfo  {journal} {Phys. Rev. C}\
  }\textbf {\bibinfo {volume} {90}},\ \bibinfo {pages} {045803} (\bibinfo
  {year} {2014})}\BibitemShut {NoStop}%
\bibitem [{\citenamefont {{Gulminelli}}\ and\ \citenamefont
  {{Raduta}}(2015)}]{Gulminelli2015}%
  \BibitemOpen
  \bibfield  {author} {\bibinfo {author} {\bibfnamefont {F.}~\bibnamefont
  {{Gulminelli}}}\ and\ \bibinfo {author} {\bibfnamefont {A.~R.}\ \bibnamefont
  {{Raduta}}},\ }\href {\doibase 10.1103/PhysRevC.92.055803} {\bibfield
  {journal} {\bibinfo  {journal} {\prc}\ }\textbf {\bibinfo {volume} {92}},\
  \bibinfo {eid} {055803} (\bibinfo {year} {2015})},\ \Eprint
  {http://arxiv.org/abs/1504.04493} {arXiv:1504.04493 [nucl-th]} \BibitemShut
  {NoStop}%
\bibitem [{\citenamefont {{Haensel}}\ \emph {et~al.}(1989)\citenamefont
  {{Haensel}}, \citenamefont {{Zdunik}},\ and\ \citenamefont
  {{Dobaczewski}}}]{HZD89}%
  \BibitemOpen
  \bibfield  {author} {\bibinfo {author} {\bibfnamefont {P.}~\bibnamefont
  {{Haensel}}}, \bibinfo {author} {\bibfnamefont {J.~L.}\ \bibnamefont
  {{Zdunik}}}, \ and\ \bibinfo {author} {\bibfnamefont {J.}~\bibnamefont
  {{Dobaczewski}}},\ }\href@noop {} {\bibfield  {journal} {\bibinfo  {journal}
  {Astron. Astrophys.}\ }\textbf {\bibinfo {volume} {222}},\ \bibinfo {pages}
  {353} (\bibinfo {year} {1989})}\BibitemShut {NoStop}%
\bibitem [{\citenamefont {Negele}\ and\ \citenamefont
  {Vautherin}(1973)}]{Negele73}%
  \BibitemOpen
  \bibfield  {author} {\bibinfo {author} {\bibfnamefont {J.~W.}\ \bibnamefont
  {Negele}}\ and\ \bibinfo {author} {\bibfnamefont {D.}~\bibnamefont
  {Vautherin}},\ }\href@noop {} {\bibfield  {journal} {\bibinfo  {journal}
  {Nucl. Phy.}\ }\textbf {\bibinfo {volume} {A207}},\ \bibinfo {pages} {298}
  (\bibinfo {year} {1973})}\BibitemShut {NoStop}%
\bibitem [{\citenamefont {Weinberg}(1972)}]{Weinberg72}%
  \BibitemOpen
  \bibfield  {author} {\bibinfo {author} {\bibfnamefont {S.}~\bibnamefont
  {Weinberg}},\ }\href@noop {} {\emph {\bibinfo {title} {Gravitation and
  Cosmology}}}\ (\bibinfo  {publisher} {Wiley, New York},\ \bibinfo {year}
  {1972})\BibitemShut {NoStop}%
\bibitem [{\citenamefont {Antoniadis}\ and\ \citenamefont {{\it et.
  al}}(2013)}]{Antoniadis13}%
  \BibitemOpen
  \bibfield  {author} {\bibinfo {author} {\bibfnamefont {J.}~\bibnamefont
  {Antoniadis}}\ and\ \bibinfo {author} {\bibnamefont {{\it et. al}}},\
  }\href@noop {} {\bibfield  {journal} {\bibinfo  {journal} {Science}\ }\textbf
  {\bibinfo {volume} {340}},\ \bibinfo {pages} {448} (\bibinfo {year}
  {2013})}\BibitemShut {NoStop}%
\bibitem [{\citenamefont {Demorest}\ \emph {et~al.}(2010)\citenamefont
  {Demorest}, \citenamefont {Pennucci}, \citenamefont {Ransom}, \citenamefont
  {Roberts},\ and\ \citenamefont {Hessels}}]{Demorest10}%
  \BibitemOpen
  \bibfield  {author} {\bibinfo {author} {\bibfnamefont {P.~B.}\ \bibnamefont
  {Demorest}}, \bibinfo {author} {\bibfnamefont {T.}~\bibnamefont {Pennucci}},
  \bibinfo {author} {\bibfnamefont {S.~M.}\ \bibnamefont {Ransom}}, \bibinfo
  {author} {\bibfnamefont {M.~S.~E.}\ \bibnamefont {Roberts}}, \ and\ \bibinfo
  {author} {\bibfnamefont {J.~W.~T.}\ \bibnamefont {Hessels}},\ }\href@noop {}
  {\bibfield  {journal} {\bibinfo  {journal} {Nature}\ }\textbf {\bibinfo
  {volume} {467}},\ \bibinfo {pages} {1081} (\bibinfo {year}
  {2010})}\BibitemShut {NoStop}%
\bibitem [{\citenamefont {Fonseca}\ \emph {et~al.}(2016)\citenamefont {Fonseca}
  \emph {et~al.}}]{Fonseca16}%
  \BibitemOpen
  \bibfield  {author} {\bibinfo {author} {\bibfnamefont {E.}~\bibnamefont
  {Fonseca}} \emph {et~al.},\ }\href@noop {} {\  (\bibinfo {year} {2016})},\
  \Eprint {http://arxiv.org/abs/1603.00545} {arXiv:1603.00545 [astro-ph.HE]}
  \BibitemShut {NoStop}%
%%CITATION = ARXIV:1603.00545;%%
\bibitem [{\citenamefont {Fantina}\ \emph {et~al.}(2014)\citenamefont
  {Fantina}, \citenamefont {Chamel}, \citenamefont {Pearson},\ and\
  \citenamefont {Goriely}}]{Fantina14}%
  \BibitemOpen
  \bibfield  {author} {\bibinfo {author} {\bibfnamefont {A.~F.}\ \bibnamefont
  {Fantina}}, \bibinfo {author} {\bibfnamefont {N.}~\bibnamefont {Chamel}},
  \bibinfo {author} {\bibfnamefont {J.~M.}\ \bibnamefont {Pearson}}, \ and\
  \bibinfo {author} {\bibfnamefont {S.}~\bibnamefont {Goriely}},\ }\href@noop
  {} {\bibfield  {journal} {\bibinfo  {journal} {EPJ Web of Conferences}\
  }\textbf {\bibinfo {volume} {66}},\ \bibinfo {pages} {07005} (\bibinfo {year}
  {2014})}\BibitemShut {NoStop}%
\bibitem [{\citenamefont {{Pearson}}\ \emph {et~al.}(2014)\citenamefont
  {{Pearson}}, \citenamefont {{Chamel}}, \citenamefont {{Fantina}},\ and\
  \citenamefont {{Goriely}}}]{Pearson2014}%
  \BibitemOpen
  \bibfield  {author} {\bibinfo {author} {\bibfnamefont {J.~M.}\ \bibnamefont
  {{Pearson}}}, \bibinfo {author} {\bibfnamefont {N.}~\bibnamefont {{Chamel}}},
  \bibinfo {author} {\bibfnamefont {A.~F.}\ \bibnamefont {{Fantina}}}, \ and\
  \bibinfo {author} {\bibfnamefont {S.}~\bibnamefont {{Goriely}}},\ }\href
  {\doibase 10.1140/epja/i2014-14043-8} {\bibfield  {journal} {\bibinfo
  {journal} {European Physical Journal A}\ }\textbf {\bibinfo {volume} {50}},\
  \bibinfo {eid} {43} (\bibinfo {year} {2014})},\ \Eprint
  {http://arxiv.org/abs/1309.2783} {arXiv:1309.2783 [nucl-th]} \BibitemShut
  {NoStop}%
\bibitem [{\citenamefont {Brandt}(1997)}]{Brandt97}%
  \BibitemOpen
  \bibfield  {author} {\bibinfo {author} {\bibfnamefont {S.}~\bibnamefont
  {Brandt}},\ }\href@noop {} {\emph {\bibinfo {title} {Statistical and
  Computational Methods in Data Analysis}}}\ (\bibinfo  {publisher} {Springer,
  New York, 3rd English edition},\ \bibinfo {year} {1997})\BibitemShut
  {NoStop}%
\bibitem [{\citenamefont {Lattimer}\ and\ \citenamefont
  {Prakash}(2001)}]{Lattimer01}%
  \BibitemOpen
  \bibfield  {author} {\bibinfo {author} {\bibfnamefont {J.~M.}\ \bibnamefont
  {Lattimer}}\ and\ \bibinfo {author} {\bibfnamefont {M.}~\bibnamefont
  {Prakash}},\ }\href@noop {} {\bibfield  {journal} {\bibinfo  {journal}
  {Astrophys. J.}\ }\textbf {\bibinfo {volume} {550}},\ \bibinfo {pages} {426}
  (\bibinfo {year} {2001})}\BibitemShut {NoStop}%
\bibitem [{\citenamefont {De}\ \emph {et~al.}(2015)\citenamefont {De},
  \citenamefont {Samaddar},\ and\ \citenamefont {Agrawal}}]{De15}%
  \BibitemOpen
  \bibfield  {author} {\bibinfo {author} {\bibfnamefont {J.~N.}\ \bibnamefont
  {De}}, \bibinfo {author} {\bibfnamefont {S.~K.}\ \bibnamefont {Samaddar}}, \
  and\ \bibinfo {author} {\bibfnamefont {B.~K.}\ \bibnamefont {Agrawal}},\
  }\href@noop {} {\bibfield  {journal} {\bibinfo  {journal} {Phys. Rev.}\
  }\textbf {\bibinfo {volume} {C92}},\ \bibinfo {pages} {014304} (\bibinfo
  {year} {2015})}\BibitemShut {NoStop}%
\bibitem [{\citenamefont {Trippa}\ \emph {et~al.}(2008)\citenamefont {Trippa},
  \citenamefont {Col{\`o}},\ and\ \citenamefont {Vigezzi}}]{Trippa08}%
  \BibitemOpen
  \bibfield  {author} {\bibinfo {author} {\bibfnamefont {L.}~\bibnamefont
  {Trippa}}, \bibinfo {author} {\bibfnamefont {G.}~\bibnamefont {Col{\`o}}}, \
  and\ \bibinfo {author} {\bibfnamefont {E.}~\bibnamefont {Vigezzi}},\
  }\href@noop {} {\bibfield  {journal} {\bibinfo  {journal} {Phys. Rev. C}\
  }\textbf {\bibinfo {volume} {77}},\ \bibinfo {pages} {061304(R)} (\bibinfo
  {year} {2008})}\BibitemShut {NoStop}%
\bibitem [{\citenamefont {Roca-Maza}\ \emph
  {et~al.}(2013{\natexlab{a}})\citenamefont {Roca-Maza}, \citenamefont
  {Brenna}, \citenamefont {Col{\`o}}, \citenamefont {Centelles}, \citenamefont
  {Vi{\~n}as}, \citenamefont {Agrawal}, \citenamefont {Paar}, \citenamefont
  {Vretenar},\ and\ \citenamefont {Piekarewicz}}]{Roca-Maza13a}%
  \BibitemOpen
  \bibfield  {author} {\bibinfo {author} {\bibfnamefont {X.}~\bibnamefont
  {Roca-Maza}}, \bibinfo {author} {\bibfnamefont {M.}~\bibnamefont {Brenna}},
  \bibinfo {author} {\bibfnamefont {G.}~\bibnamefont {Col{\`o}}}, \bibinfo
  {author} {\bibfnamefont {M.}~\bibnamefont {Centelles}}, \bibinfo {author}
  {\bibfnamefont {X.}~\bibnamefont {Vi{\~n}as}}, \bibinfo {author}
  {\bibfnamefont {B.~K.}\ \bibnamefont {Agrawal}}, \bibinfo {author}
  {\bibfnamefont {N.}~\bibnamefont {Paar}}, \bibinfo {author} {\bibfnamefont
  {D.}~\bibnamefont {Vretenar}}, \ and\ \bibinfo {author} {\bibfnamefont
  {J.}~\bibnamefont {Piekarewicz}},\ }\href@noop {} {\bibfield  {journal}
  {\bibinfo  {journal} {Phys. Rev. C}\ }\textbf {\bibinfo {volume} {88}},\
  \bibinfo {pages} {024316} (\bibinfo {year} {2013}{\natexlab{a}})}\BibitemShut
  {NoStop}%
\bibitem [{\citenamefont {Roca-Maza}\ \emph
  {et~al.}(2013{\natexlab{b}})\citenamefont {Roca-Maza}, \citenamefont
  {Brenna}, \citenamefont {Agrawal}, \citenamefont {Bortignon}, \citenamefont
  {Col{\`o}}, \citenamefont {Cao}, \citenamefont {Paar},\ and\ \citenamefont
  {Vretenar}}]{Roca-maza13}%
  \BibitemOpen
  \bibfield  {author} {\bibinfo {author} {\bibfnamefont {X.}~\bibnamefont
  {Roca-Maza}}, \bibinfo {author} {\bibfnamefont {M.}~\bibnamefont {Brenna}},
  \bibinfo {author} {\bibfnamefont {B.~K.}\ \bibnamefont {Agrawal}}, \bibinfo
  {author} {\bibfnamefont {P.~F.}\ \bibnamefont {Bortignon}}, \bibinfo {author}
  {\bibfnamefont {G.}~\bibnamefont {Col{\`o}}}, \bibinfo {author}
  {\bibfnamefont {L.-G.}\ \bibnamefont {Cao}}, \bibinfo {author} {\bibfnamefont
  {N.}~\bibnamefont {Paar}}, \ and\ \bibinfo {author} {\bibfnamefont
  {D.}~\bibnamefont {Vretenar}},\ }\href@noop {} {\bibfield  {journal}
  {\bibinfo  {journal} {Phys. Rev. C}\ }\textbf {\bibinfo {volume} {87}},\
  \bibinfo {pages} {034301} (\bibinfo {year} {2013}{\natexlab{b}})}\BibitemShut
  {NoStop}%
\bibitem [{\citenamefont {Carbone}\ \emph {et~al.}(2010)\citenamefont
  {Carbone}, \citenamefont {Col{\`o}}, \citenamefont {Bracco}, \citenamefont
  {Cao}, \citenamefont {Bortignon}, \citenamefont {Camera},\ and\ \citenamefont
  {Wieland}}]{Carbone10}%
  \BibitemOpen
  \bibfield  {author} {\bibinfo {author} {\bibfnamefont {A.}~\bibnamefont
  {Carbone}}, \bibinfo {author} {\bibfnamefont {G.}~\bibnamefont {Col{\`o}}},
  \bibinfo {author} {\bibfnamefont {A.}~\bibnamefont {Bracco}}, \bibinfo
  {author} {\bibfnamefont {L.-G.}\ \bibnamefont {Cao}}, \bibinfo {author}
  {\bibfnamefont {P.~F.}\ \bibnamefont {Bortignon}}, \bibinfo {author}
  {\bibfnamefont {F.}~\bibnamefont {Camera}}, \ and\ \bibinfo {author}
  {\bibfnamefont {O.}~\bibnamefont {Wieland}},\ }\href@noop {} {\bibfield
  {journal} {\bibinfo  {journal} {Phys. Rev. C}\ }\textbf {\bibinfo {volume}
  {81}},\ \bibinfo {pages} {041301(R)} (\bibinfo {year} {2010})}\BibitemShut
  {NoStop}%
\bibitem [{\citenamefont {Chen}(2011)}]{Chen11}%
  \BibitemOpen
  \bibfield  {author} {\bibinfo {author} {\bibfnamefont {L.-W.}\ \bibnamefont
  {Chen}},\ }\href@noop {} {\bibfield  {journal} {\bibinfo  {journal} {Phys.
  Rev. C}\ }\textbf {\bibinfo {volume} {83}},\ \bibinfo {pages} {044308}
  (\bibinfo {year} {2011})}\BibitemShut {NoStop}%
\bibitem [{\citenamefont {Roca-Maza}\ \emph {et~al.}(2011)\citenamefont
  {Roca-Maza}, \citenamefont {Vi˜nas}, \citenamefont {Centelles},
  \citenamefont {Ring},\ and\ \citenamefont {Schuck}}]{Roca-Maza11}%
  \BibitemOpen
  \bibfield  {author} {\bibinfo {author} {\bibfnamefont {X.}~\bibnamefont
  {Roca-Maza}}, \bibinfo {author} {\bibfnamefont {X.}~\bibnamefont {Vi˜nas}},
  \bibinfo {author} {\bibfnamefont {M.}~\bibnamefont {Centelles}}, \bibinfo
  {author} {\bibfnamefont {P.}~\bibnamefont {Ring}}, \ and\ \bibinfo {author}
  {\bibfnamefont {P.}~\bibnamefont {Schuck}},\ }\href@noop {} {\bibfield
  {journal} {\bibinfo  {journal} {Phys. Rev. C}\ }\textbf {\bibinfo {volume}
  {84}},\ \bibinfo {pages} {054309} (\bibinfo {year} {2011})}\BibitemShut
  {NoStop}%
\bibitem [{\citenamefont {Nik{\v s}i{\'c}}\ \emph {et~al.}(2008)\citenamefont
  {Nik{\v s}i{\'c}}, \citenamefont {Vretenar},\ and\ \citenamefont
  {Ring}}]{Niksic08}%
  \BibitemOpen
  \bibfield  {author} {\bibinfo {author} {\bibfnamefont {T.}~\bibnamefont
  {Nik{\v s}i{\'c}}}, \bibinfo {author} {\bibfnamefont {D.}~\bibnamefont
  {Vretenar}}, \ and\ \bibinfo {author} {\bibfnamefont {P.}~\bibnamefont
  {Ring}},\ }\href@noop {} {\bibfield  {journal} {\bibinfo  {journal} {Phys.
  Rev. C}\ }\textbf {\bibinfo {volume} {78}},\ \bibinfo {pages} {034318}
  (\bibinfo {year} {2008})}\BibitemShut {NoStop}%
\bibitem [{\citenamefont {Zhao}\ \emph {et~al.}(2010)\citenamefont {Zhao},
  \citenamefont {Lin}, \citenamefont {Yao},\ and\ \citenamefont
  {Meng}}]{Zhao10}%
  \BibitemOpen
  \bibfield  {author} {\bibinfo {author} {\bibfnamefont {P.}~\bibnamefont
  {Zhao}}, \bibinfo {author} {\bibfnamefont {Z.~P.}\ \bibnamefont {Lin}},
  \bibinfo {author} {\bibfnamefont {J.~M.}\ \bibnamefont {Yao}}, \ and\
  \bibinfo {author} {\bibfnamefont {J.}~\bibnamefont {Meng}},\ }\href@noop {}
  {\bibfield  {journal} {\bibinfo  {journal} {Phys. Rev. C}\ }\textbf {\bibinfo
  {volume} {82}},\ \bibinfo {pages} {054319} (\bibinfo {year}
  {2010})}\BibitemShut {NoStop}%
\bibitem [{\citenamefont {Tsang}\ \emph {et~al.}(2012)\citenamefont {Tsang},
  \citenamefont {Stone}, \citenamefont {Camera}, \citenamefont {Danielewicz},
  \citenamefont {Gandolfi}, \citenamefont {Hebeler}, \citenamefont {Horowitz},
  \citenamefont {Lee}, \citenamefont {Lynch}, \citenamefont {Kohley},
  \citenamefont {Lemmon}, \citenamefont {M\"oller}, \citenamefont {Murakami},
  \citenamefont {Riordan}, \citenamefont {Roca-Maza}, \citenamefont
  {Sammarruca}, \citenamefont {Steiner}, \citenamefont {Vida\~na},\ and\
  \citenamefont {Yennello}}]{Tsang12}%
  \BibitemOpen
  \bibfield  {author} {\bibinfo {author} {\bibfnamefont {M.~B.}\ \bibnamefont
  {Tsang}}, \bibinfo {author} {\bibfnamefont {J.~R.}\ \bibnamefont {Stone}},
  \bibinfo {author} {\bibfnamefont {F.}~\bibnamefont {Camera}}, \bibinfo
  {author} {\bibfnamefont {P.}~\bibnamefont {Danielewicz}}, \bibinfo {author}
  {\bibfnamefont {S.}~\bibnamefont {Gandolfi}}, \bibinfo {author}
  {\bibfnamefont {K.}~\bibnamefont {Hebeler}}, \bibinfo {author} {\bibfnamefont
  {C.~J.}\ \bibnamefont {Horowitz}}, \bibinfo {author} {\bibfnamefont
  {J.}~\bibnamefont {Lee}}, \bibinfo {author} {\bibfnamefont {W.~G.}\
  \bibnamefont {Lynch}}, \bibinfo {author} {\bibfnamefont {Z.}~\bibnamefont
  {Kohley}}, \bibinfo {author} {\bibfnamefont {R.}~\bibnamefont {Lemmon}},
  \bibinfo {author} {\bibfnamefont {P.}~\bibnamefont {M\"oller}}, \bibinfo
  {author} {\bibfnamefont {T.}~\bibnamefont {Murakami}}, \bibinfo {author}
  {\bibfnamefont {S.}~\bibnamefont {Riordan}}, \bibinfo {author} {\bibfnamefont
  {X.}~\bibnamefont {Roca-Maza}}, \bibinfo {author} {\bibfnamefont
  {F.}~\bibnamefont {Sammarruca}}, \bibinfo {author} {\bibfnamefont {A.~W.}\
  \bibnamefont {Steiner}}, \bibinfo {author} {\bibfnamefont {I.}~\bibnamefont
  {Vida\~na}}, \ and\ \bibinfo {author} {\bibfnamefont {S.~J.}\ \bibnamefont
  {Yennello}},\ }\href@noop {} {\bibfield  {journal} {\bibinfo  {journal}
  {Phys. Rev. C}\ }\textbf {\bibinfo {volume} {86}},\ \bibinfo {pages} {015803}
  (\bibinfo {year} {2012})}\BibitemShut {NoStop}%
\bibitem [{\citenamefont {Khan}\ \emph {et~al.}(2012)\citenamefont {Khan},
  \citenamefont {Margueron},\ and\ \citenamefont {Vida{\~n}a}}]{Khan12}%
  \BibitemOpen
  \bibfield  {author} {\bibinfo {author} {\bibfnamefont {E.}~\bibnamefont
  {Khan}}, \bibinfo {author} {\bibfnamefont {J.}~\bibnamefont {Margueron}}, \
  and\ \bibinfo {author} {\bibfnamefont {I.}~\bibnamefont {Vida{\~n}a}},\
  }\href@noop {} {\bibfield  {journal} {\bibinfo  {journal} {Phys. Rev. Lett.}\
  }\textbf {\bibinfo {volume} {109}},\ \bibinfo {pages} {092501} (\bibinfo
  {year} {2012})}\BibitemShut {NoStop}%
\bibitem [{\citenamefont {Khan}\ and\ \citenamefont
  {Margueron}(2013)}]{khan13}%
  \BibitemOpen
  \bibfield  {author} {\bibinfo {author} {\bibfnamefont {E.}~\bibnamefont
  {Khan}}\ and\ \bibinfo {author} {\bibfnamefont {J.}~\bibnamefont
  {Margueron}},\ }\href@noop {} {\bibfield  {journal} {\bibinfo  {journal}
  {Phys. Rev. C}\ }\textbf {\bibinfo {volume} {88}},\ \bibinfo {pages} {034319}
  (\bibinfo {year} {2013})}\BibitemShut {NoStop}%
\bibitem [{\citenamefont {Steiner}\ \emph {et~al.}(2016)\citenamefont
  {Steiner}, \citenamefont {Lattimer},\ and\ \citenamefont
  {Brown}}]{Steiner16}%
  \BibitemOpen
  \bibfield  {author} {\bibinfo {author} {\bibfnamefont {A.~W.}\ \bibnamefont
  {Steiner}}, \bibinfo {author} {\bibfnamefont {J.~M.}\ \bibnamefont
  {Lattimer}}, \ and\ \bibinfo {author} {\bibfnamefont {E.~F.}\ \bibnamefont
  {Brown}},\ }\href@noop {} {\bibfield  {journal} {\bibinfo  {journal} {Eur.
  Phys. J.}\ }\textbf {\bibinfo {volume} {A52}},\ \bibinfo {pages} {18}
  (\bibinfo {year} {2016})}\BibitemShut {NoStop}%
\end{thebibliography}

%

\end{document}